\documentclass[prb,twocolumn,10pt,amsmath,amsfonts,amssymb]{revtex4-1}
\newcommand{\expect}[1]{\langle #1 \rangle}
\newcommand{\overlap}[2]{\langle #1 | #2 \rangle}
\newcommand{\bra}[1]{\langle #1 |}
\newcommand{\ket}[1]{| #1 \rangle}
\newcommand{\tab}[1]{Table \ref{tab:#1}}

\usepackage{graphicx}

\begin{document}
\title{Sensitivity of the photo-physical properties of organometallic complexes to small chemical changes}
\author{A. C. Jacko}
\email{jacko@physics.uq.edu.au}
\affiliation{Centre for Organic Photonics and Electronics, School of Mathematics and Physics, The University of Queensland}
\author{B. J. Powell}
\affiliation{Centre for Organic Photonics and Electronics, School of Mathematics and Physics, The University of Queensland}
\author{Ross H. McKenzie}
\affiliation{Centre for Organic Photonics and Electronics, School of Mathematics and Physics, The University of Queensland}
\date{\today}

\begin{abstract}
We investigate an effective model Hamiltonian for organometallic complexes that are widely used in optoelectronic devices. The two most important parameters in the model are $J$, the effective exchange interaction between the $\pi$ and $\pi^*$ orbitals of the ligands, and $\varepsilon^*$, the renormalized energy gap between the highest occupied orbitals on the metal and on the ligand. We find that the degree of metal-to-ligand charge transfer (MLCT) character of the lowest triplet state is strongly dependent on the ratio $\varepsilon^*/J$. $\varepsilon^*$ is purely a property of the complex and can be changed significantly by even small variations in the complex's chemistry, such as replacing substituents on the ligands.  We find that that small changes in  $\varepsilon^*/J$ can cause large changes in the properties of the complex, including the lifetime of the triplet state and the probability of injected charges (electrons and holes) forming triplet excitations. These results give some insight into the observed large changes in the photophysical properties of organometallic complexes caused by small changes in the ligands.
\end{abstract}

\maketitle

\section{Introduction}

Organometallic complexes are being developed as optically active materials for devices such as organic light emitting diodes (OLED),\cite{friend99,forrest04,li05,lo06} and organic photovoltaics (OPV).\cite{gratzel91,hagfeldt00,dimitrakopoulos02,fernandez08}
The functionality of such complexes depends on their excited state properties. Of particular interest is the emission process, key to the function of organic light emitting diodes (OLED),
\[
\text{electron} + \text{hole} \quad \rightarrow \quad C^* \quad \rightarrow \quad C + \text{photon}
\]
where $C^*$ ($C$) denotes the complex in its excited (ground) state. The reverse process, converting a photon into electron and hole excitations, is the key process in OPV cells such as dye-sensitized solar cells (Gratzel cells).\cite{gratzel91}  To understand and control these processes (in the design of new complexes) one needs to understand the relevant excited states of the complex. 

\begin{figure}[ht]
 \centering
 \includegraphics[width=0.9\columnwidth]{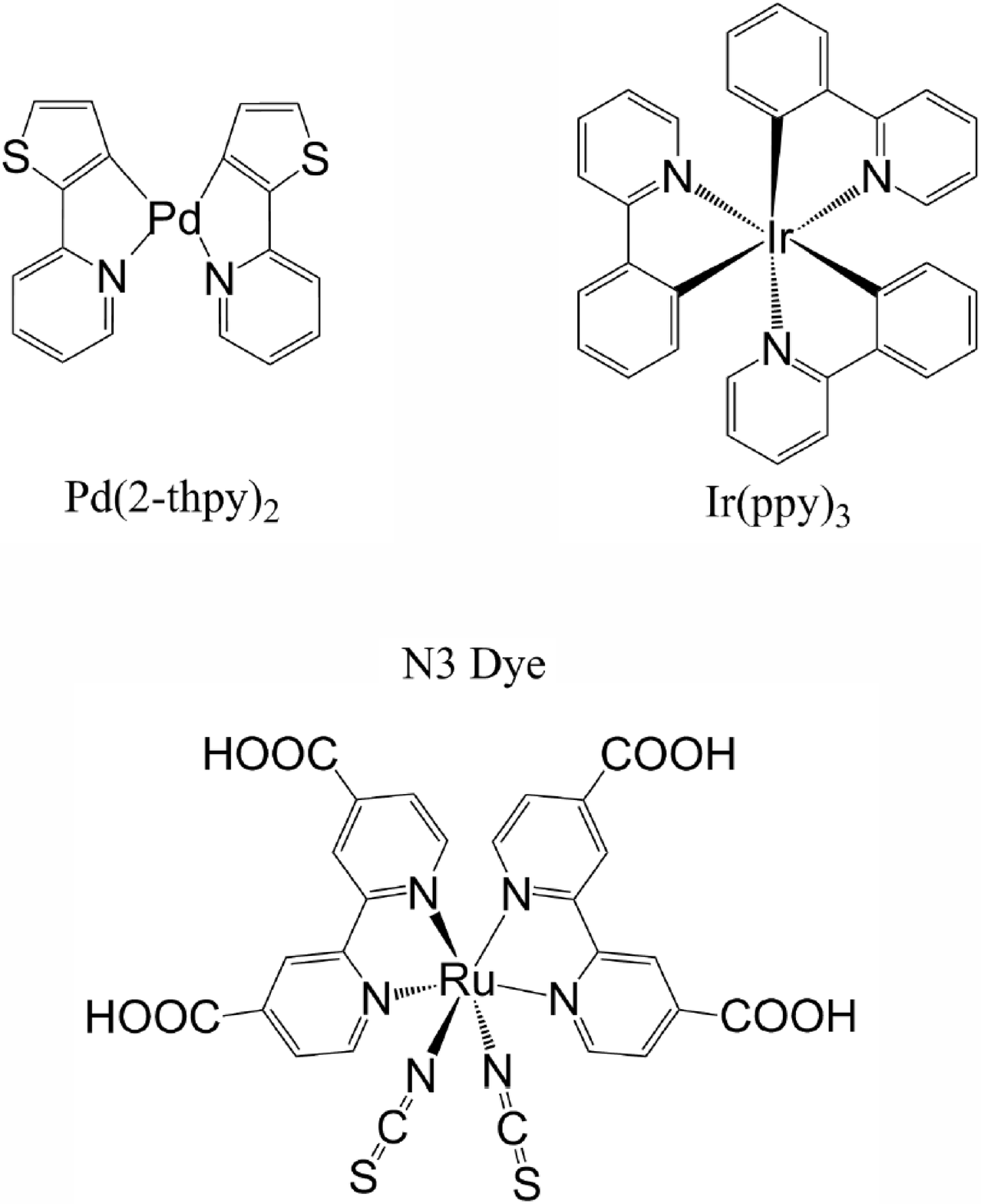}
 \caption[Examples of organometallic complexes.]{Examples of organometallic complexes for optoelectronic applications. Both Pd(thpy)$_2$ and Ir(ppy)$_3$ are the basis of many complexes used in organic light emitting diodes \cite{li05,yersin02}, while the benchmark N3 dye \cite{nazeeruddin93} is used in organic dye-sensitized photovoltaic cells.}\label{fig:complexes}
\end{figure}
Two types of excited state typically dominate the behavior of organometallic materials (such as those illustrated in Fig. \ref{fig:complexes}): ligand centered (LC) states and metal to ligand charge transfer (MLCT) states. Isolated ligands have a `bright' high energy (UV) singlet transition (associated with a $\pi \rightarrow \pi^*$ excitation, equivalent to a LC transition). This transition has a molar absorptivity of order $10^4$ mol$^{-1}$ cm$^{-1}$.\cite{yersinbook} The corresponding $\pi \rightarrow \pi^*$ triplet transition has a much lower energy and is `dark'. \cite{yersinbook} When ligands are bonded to a transition metal to form a complex new singlet transitions in the visible region are observed (with molar absorptivity of order $10^3$ mol$^{-1}$ cm$^{-1}$).\cite{yersinbook} These new features are typically attributed to MLCT transitions. The significant oscillator strength associated with MLCT singlet states arises because of the hybridization of metal orbitals with $\pi^*$ (unoccupied ligand) orbitals, known as back-bonding.\cite{hartley73,creutz69,reimers91,rusanova06}

The presence of the heavy transition metal ion produces a spin-orbit interaction which mixes singlets and triplets with several important effects.\cite{kober82}
First, this interaction allows states with dominant triplet character to decay radiatively. Second, it also allows a rapid ($\sim 50$ fs) intersystem crossing (ISC) between singlet and triplet dominated states.\cite{yoon06} In organic complexes with no transition metal ion the spin-orbit coupling is much weaker, so triplet excitations tend to decay via non-radiative decay paths. \cite{segal03,burin98,reufer05}

The character of the emitting state in organometallic complexes has attracted considerable interest and debate. It is generally agreed to be predominantly triplet (due to its long radiative lifetime, which is associated with phosphorescence). However for a variety of complexes it has been labeled variously as LC, MLCT, or as a LC-MLCT hybrid.\cite{schwarz89,colombo93,lamansky01} For example, Li \emph{et al.}\cite{li05}  considered a family of OLED complexes and claimed that the emitting state is a superposition of a singlet MLCT state and a triplet LC state. Such spin hybridization is only possible due to spin-orbit coupling. In the case of very strong spin-orbit coupling, the predominantly triplet state can have a comparable intensity to singlet emission.\cite{lamansky01}
Yersin \emph{et al.} \cite{yersin01,finkenzeller07} claimed that the amount of zero-field splitting (ZFS) in the emitting triplet reflects the amount of MLCT character in the state, and that there is ``an empirical correlation between the amount of ZFS and the compound's potential for its use as emitter material in an OLED''.\cite{finkenzeller07} It has been suggested that the key effect of increasing MLCT character is to enhance the effect of spin-orbit coupling, which in turn increases the radiative rate of triplet excited states.\cite{haneder08}

A key aspect to understanding the photophysical properties of complex molecular materials is identifying the relevant frontier molecular orbitals and their interactions with one another. This allows one to define an effective Hamiltonian which involves just a few parameters. Well known examples of this approach involve the H\"uckel, Hubbard, Heisenberg and Pariser-Parr-Pople models for conjugated polyenes.\cite{powell09book} With regard to organo-metallic complexes this approach has been applied to mixed valence binuclear systems including magnetic atoms in proteins (Hubbard and double exchange models)\cite{blondin90}, molecular magnets,\cite{nagao00} and Anderson impurity models for cobalt based valence tautomers.\cite{labute02}

Such semi-empirical approaches have significant advantages which mean that they nicely complement first principles approaches such as DFT and high level \textit{ab initio} quantum chemistry.\cite{jacko10b} First, effective Hamiltonians can help to reveal features that are common to a diverse range of materials. Second, since the number of degrees of freedom in the model is significantly less than in the actual material one does not necessarily have to make the approximations one would be forced to make if one did a complete quantum simulation of the actual material. For example, for large molecular systems one can also describe the nuclear dynamics quantum mechanically\cite{worth04} and include the effect of the environment such as the solvent.\cite{terenziani06,gilmore05} Such models can capture universal behavior that is not sensitive to microscopic details. For example, the  single impurity Anderson model can describe the Kondo effect in a diverse range of systems including magnetic impurities in metals, quantum dots in semiconductor heterostructures, carbon nanotubes, and single molecule transistors.\cite{hewsonbook,kouwenhoven01}

Here we develop and examine a model Hamiltonian which reproduces the key photo-physical properties of organometallic complexes. We apply the model to complexes in which the low energy physics is dominated by one ligand (see for example Refs. \onlinecite{rusanova06}, \onlinecite{haneder08}, \onlinecite{bomben09}). We find that there are two key parameters in describing the low energy photophysics of these complexes - $J$, the spin-exchange interaction between ligand $\pi$ and $\pi^*$ orbitals, and $\varepsilon^*$, the renormalized energy gap between the ligand HOMO and highest occupied metal orbital. We show that, through small changes to these parameters, single chemical substitutions can cause significant changes in the triplet excited state lifetime and the probability of triplet formation, causing large variations in efficiency. 
This paper is organized as follows. In Sec. II we introduce our model Hamiltonian, discussing its parameters and an appropriate basis in which to investigate its properties. In Sec. III we analyze some approximate and exact solutions of the Hamiltonian, building an understanding of the key features of the model. In Sec. IV we determine the effect of MLCT character on the radiative properties of the excited states. In Sec. V we study the probability of finding an excitation in the triplet state after charge injection. Sec. VI presents some concluding remarks.

\section{Model Hamiltonian}

In principle, one should determine an effective Hamiltonian by integrating out high energy states. However, explicitly carrying out this procedure is prohibitively expensive for all but the smallest molecules.\cite{freed83} Therefore, in order to investigate correlation effects in these organometallic complexes we reduce the size of the basis set in a way motivated by our understanding of the important physical processes. 

To this end we study the following model  Hamiltonian:
\begin{eqnarray}\label{eq:thehamiltonian11}
\hat{H}= &-& J \vec{S}_{H} \cdot \vec{S}_{L} +\sum_{\sigma=\uparrow,\downarrow} \Big\{ \varepsilon M^\dagger_{\sigma} M_{\sigma} + \Delta L^\dagger_{\sigma} L_{\sigma} \nonumber \\
&-& t^{H} \left(H^\dagger_{\sigma} M_{\sigma} + M^\dagger_{\sigma} H_{\sigma} \right) - t^{L} \left(L^\dagger_{\sigma} M_{\sigma} + M^\dagger_{\sigma} L_{\sigma} \right)  \Big\} \nonumber \\
&+& U_H n_{H\uparrow}n_{H\downarrow} + U_L n_{L\uparrow}n_{L\downarrow} + U_M n_{M\uparrow}n_{M\downarrow} \nonumber \\
&+& V_{HL} n_{H} n_{L}  +  V_{HM} n_{H} n_{M} + V_{ML}n_{M} n_{L},
\end{eqnarray}
where 
$n_H \equiv \sum_{\sigma} H^\dagger_{\sigma} H_{\sigma}$,
$n_L \equiv \sum_{\sigma} L^\dagger_{\sigma} L_{\sigma}$,
$n_M \equiv \sum_{\sigma} M^\dagger_{\sigma} M_{\sigma}$,
$\vec{S}_H = \sum_{\sigma} \sum_{\sigma'} H_\sigma^\dagger \vec{\sigma}_{\sigma \sigma'} H_\sigma'$,
$\vec{S}_L = \sum_{\sigma} \sum_{\sigma'} L_\sigma^\dagger \vec{\sigma}_{\sigma \sigma'} L_\sigma'$,
$\vec{\sigma}_{\sigma \sigma'} = \left(\hat{\sigma}^x_{\sigma, \sigma'},\hat{\sigma}^y_{\sigma, \sigma'},\hat{\sigma}^z_{\sigma, \sigma'} \right)$ where $\hat{\sigma}^d_{\sigma, \sigma'}$ is the $(\sigma, \sigma')$ element of the $d$ Pauli spin matrix, $\hat{\sigma}^d$.
In what follows it will be useful to have intuitive descriptions of the creation operators $H^{\dagger}_{\sigma}$, $L^{\dagger}_{\sigma}$, and $M^{\dagger}_{\sigma}$. Thus, we will refer to the state $H^\dagger_{\sigma}|vac\rangle$ (where $|vac\rangle$ is the state with $\langle n_H\rangle=\langle n_M\rangle=\langle n_L\rangle=0$) as the `HOMO level of the ligand', call $L^\dagger_{\sigma}|vac\rangle$ the `LUMO level of the ligand', and call $M^\dagger_{\sigma}|vac\rangle$ the `metal orbital'. It is important to stress, however, that the level $H^\dagger_{\sigma}|vac\rangle$ ($L^\dagger_{\sigma}|vac\rangle$) may well be very different from the highest occupied (lowest unoccupied) molecular orbital that one would find in a Hartree-Fock calculation.  Similarly, $M^\dagger_{\sigma}|vac\rangle$ is likely to be very different from the atomic $d$ orbital in the isolated transition metal.\cite{powell09book} As such one cannot describe the coefficients of the Hamiltonian in terms of one electron orbitals like $U_H = \iint dr_1 dr_2 \phi_{H \uparrow}^*(1) \phi_{H \uparrow}(1) \frac{e^2}{|r_1 - r_2 |} \phi_{H \downarrow}^*(2) \phi_{H \downarrow}(2)$. 

Indeed, it is not possible, in general, to associate any static `orbital distribution' with any of these states as orbital relaxation will occur when the physical state of the complex changes.
Rather, orbital relaxation (and higher order processes) are captured in the renormalization of the effective parameters of the model.\cite{powell09book,scriven09} However, one expects that the model states correspond rather more closely to ones intuition of how a HOMO, LUMO, and metal orbital behave than the actual Hartree-Fock orbitals do. Thus, this may provide new insights to the problems at hand.\cite{jacko10b}

One might intuitively interpret the parameters as follows: $J$ is the effective exchange interaction between the ligand HOMO and the ligand LUMO, $\varepsilon$ is the effective difference in the energies of the metal orbital and the ligand HOMO, $\Delta$ is the effective HOMO-LUMO gap, $t^{H}$ $(t^{L})$ is the effective hopping amplitude between the metal orbital and the ligand HOMO (LUMO), $U_i$ is the effective Coulomb repulsion between two electrons in orbital $i$ (where $i \in \{H, L, M \}$), and $V_{ij}$ is the effective Coulomb repulsion between an electron in orbital $i$ and another in orbital $j$. This interpretation would seem to be equivalent to making an INDO (incomplete neglect of differential overlap) approximation\cite{fulde} after the physically motivated basis set reduction. However, we do not now make the Hartree-Fock approximation  made in INDO calculations. Note that these parameters are renormalised from the values that would be found from direct computation. It is known that the parameter values in small organic molecules, like the ligands considered here (cf. Fig. \ref{fig:complexes}) can be significantly changed due to this renormalization.\cite{scriven09,gunnarssonbook,brocks04,canocortes07,scriven09B} Indeed, we will not attempt to calculate the values of the parameters here. Instead we construct a semi-empirical theory by comparing our model to experimental data in Appendix \ref{App_param}.

In general, several ligands and metal orbitals could be involved in the low energy physics of the organometallic complexes we are investigating. For simplicity, here we focus on complexes in which one ligand dominates the low energy physics (for example Ru(NH$_3$)$_2$Cl$_2$(bqdi) (Ref. \onlinecite{rusanova06}), Ru(dcbpy)(bpy)$_2$ (Ref. \onlinecite{bomben09}), or Pt(cnpmic) (Ref. \onlinecite{haneder08})), interacting with only one fully occupied metal orbital, and leave the discussion of models with more metal and ligand sites for a later publication. In the non-interacting ground state $\langle n_H\rangle=\langle n_M\rangle=2$ and $\langle n_L\rangle=0$.  Thus we consider a three site model with four electrons.

\subsection*{Basis States}

One expects that the LUMO will be much higher in energy than the metal state ($\Delta-\varepsilon \gg t^{L}$, see Appendix). Therefore, the effect of $t^{L}$ on the eigenstates will be small compared to the other interactions. As such $n_L$ is almost a quantum number so it is convenient to work in a basis of states which are eigenstates of $n_L$.

We define a `reference state' for this model
\begin{equation}
\ket{0} \equiv H^\dagger_\uparrow H^\dagger_\downarrow M^\dagger_\uparrow M^\dagger_\downarrow \ket{vac} \nonumber
\end{equation}
the $n_L = 0$ state, with energy $E_0 = \bra{0}\hat{H}\ket{0} = 2 \varepsilon + U_H + U_M + 4 V_{HM}$, and where $\ket{vac}$ is the vacuum state.

We will also choose our basis states to be eigenstates of the exchange interaction, $J$, that is, singlets and triplets. We define the $n_L=1$ metal to ligand charge transfer (MLCT) triplet states as
\begin{eqnarray}                            
\ket{^{3}MLCT^1} &\equiv& \Big\{L^\dagger_\uparrow M_\downarrow \ket{0},\, L^\dagger_\downarrow M_\uparrow \ket{0},\, 
\nonumber \\ & & \frac{1}{\sqrt{2}}(L^\dagger_\uparrow M_\uparrow - L^\dagger_\downarrow M_\downarrow) \ket{0} \Big\},  \nonumber
\end{eqnarray}
and the ligand centered (LC) triplet states as
\[
\ket{^{3}LC^1} \equiv \left\{L^\dagger_\uparrow H_\downarrow \ket{0},\, L^\dagger_\downarrow H_\uparrow \ket{0},\, \frac{1}{\sqrt{2}}(L^\dagger_\uparrow H_\uparrow - L^\dagger_\downarrow H_\downarrow) \ket{0} \right\},
\]
where the prefix $^{3}$ indicates the spin degeneracy and the suffix $^{1}$ is the value of $n_L$ in these states (since these are triplets between an electron and a hole in the reference state, the $S^z = 0$ states appear to have the opposite phase relation to usual).
Similarly, the MLCT singlet is
\[
\ket{^{1}MLCT^1} \equiv  \frac{1}{\sqrt{2}}(L^\dagger_\uparrow M_\uparrow + L^\dagger_\downarrow M_\downarrow) \ket{0},
\]
and the LC singlet is
\[
\ket{^{1}LC^1} \equiv  \frac{1}{\sqrt{2}}(L^\dagger_\uparrow H_\uparrow + L^\dagger_\downarrow H_\downarrow) \ket{0}.
\]
When we go to the $n_L = 2$ states we have three singlets, the MLCT
\[
\ket{^{1}MLCT^2} \equiv L^\dagger_\uparrow L^\dagger_\downarrow M_\downarrow M_\uparrow \ket{0},
\]
the LC
\[
\ket{^{1}LC^2} \equiv L^\dagger_\uparrow L^\dagger_\downarrow H_\downarrow H_\uparrow \ket{0},
\]
and the metal-HOMO (MH) singlet
\[
\ket{^{1}MH^2} \equiv \frac{1}{\sqrt{2}}L^\dagger_\uparrow L^\dagger_\downarrow ( H_\uparrow M_\downarrow - H_\downarrow M_\uparrow) \ket{0},
\]
and only one kind of triplet,
\begin{eqnarray}
\ket{^{3}MH^2} &\equiv& \Big \{L^\dagger_\uparrow L^\dagger_\downarrow H_\downarrow M_\downarrow \ket{0} ,\, L^\dagger_\uparrow L^\dagger_\downarrow H_\uparrow M_\uparrow \ket{0} , \nonumber \\ & &  \frac{1}{\sqrt{2}}L^\dagger_\uparrow L^\dagger_\downarrow ( H_\downarrow M_\uparrow + H_\uparrow M_\downarrow) \ket{0} \Big\}. \nonumber
\end{eqnarray}

When examining the results of our model, it will be helpful to consider numerical results for a particular set of parameter values. In the Appendix we estimate parameter values relevant to complexes of interest for optoelectronic applications (such as those in Fig. \ref{fig:complexes}). We use the following parameter set as a typical example of these values: $J = 1$ eV, $\Delta = 3$ eV,  $t^H = 0.1$ eV,  $t^L = 0.1$ eV,  $U_M=U_H=U_L=U= 3$ eV,  $V_{HL} = V_{HM} = V_{LM} = V= 3$ eV (note that the approximation here that $U_H = V_{HL}$ is physically reasonable as the $H^\dagger|vac\rangle$ and $L^\dagger|vac\rangle$ states are delocalised and have a large spatial overlap; a detailed derivation of this result is given in the Appendix). It is not possible to estimate $\varepsilon$ (the energy gap between the HOMO and the metal orbital) from what is known experimentally about the ligand as it is, intrinsically, a property of the complex. Indeed, we will show below that many of the important properties of the complex depend sensitively on $\varepsilon$. For consistency and concreteness  we take  $\varepsilon=0.5$ eV when we wish to investigate how the properties of the complex vary with another parameter. In the Supplementary Information we show that our main conclusions are robust to variations of these parameters in physically reasonable ranges and present calculations for many alternative parameter sets. 

\begin{figure}
	\centering
		\includegraphics[width=0.9\columnwidth]{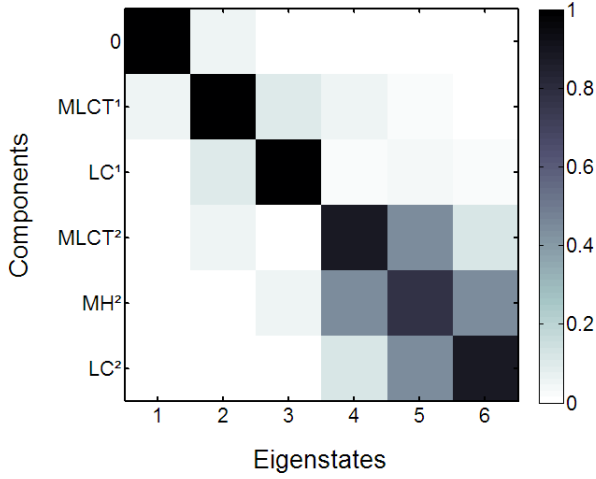}
	\caption{The eigenstates of the singlet part of the model Hamiltonian, Eq. \ref{eq:thehamiltonian11}, from lowest energy (1) to highest (6), in terms of singlet basis states which are eigenstates of $n_L$, using the example parameter set  $J = 1$ eV, $\Delta = 3$ eV,  $\varepsilon =0.25$ eV,  $t^H = 0.1$ eV,  $t^L = 0.1$ eV,  $U_M=U_H=U_L=U= 3$ eV,  $V_{HL} = V_{HM} = V_{LM} = V= 3$ eV (see Appendix for more details). It is clear that for this set of parameters the eigenstates are predominantly one basis state, and that the coupling between blocks of different $n_L$ is a small effect. These results hold for physically reasonable parameter values. See Supp. Info. for the equivalent figure at different parameter values.\cite{suppinfo} Figs. S1-S6 in the Supp. Info. show that this picture does not break down except near (avoided) level crossings, and that the first excited state has a large overlap with the MLCT$^1$ basis state for a wide range of parameters around our typical set.}
	\label{fig:eigenstates}
\end{figure}
Fig. \ref{fig:eigenstates} shows that for this typical set of parameters that $n_L$ is indeed almost a quantum number for the singlet states, particularly in the low energy singlets.

It is useful to write the Hamiltonian matrix in the basis defined above. We can write the singlet sector of the Hamiltonian in terms of blocks of constant $n_L$, $\hat{H}^{n_L}_S$, and the matrices that couple them $\hat{T}^{n_L,n_L'}_S$,
\begin{equation}
\hat{H}_S \equiv \left(\begin{array}{ccc}
	\hat{H}^0 & \hat{T}^{0,1S}_S & \hat{T}^{0,2}_S \\
	\hat{T}^{1,0}_S & \hat{H}^{1}_{S} & \hat{T}^{1,2}_S\\
	\hat{T}^{2,0}_S & \hat{T}^{2,1}_S & \hat{H}^2_S 
\end{array}\right),\label{eq:HS}
\end{equation}
and similarly for the triplet sector of the Hamiltonian,
\begin{equation}
\hat{H}_T \equiv \left(\begin{array}{cc}
	 \hat{H}^{1}_T & \hat{T}^{1,2}_T \\
	 \hat{T}^{2,1}_T & \hat{H}^2_T
\end{array}\right),\label{eq:HT}
\end{equation}
where $\hat{T}^{n_L,n_L'*}_S = \hat{T}^{n_L',n_L}_S$.
\begin{widetext}
The only $n_L=0$ basis state is the reference state, $\hat{H}^0 = (2 \varepsilon + U_H+U_M + 4 V_{HM})$. The singlet $n_L=1$ block is
\begin{equation}\label{eq:HS1}
\hat{H}^1_S=\bordermatrix{ & \ket{^{1}MLCT^1} & \ket{^{1}LC^1}  \cr
& \Delta + \varepsilon + U_H + 2V_{HL} + 2 V_{HM} + V_{LM}& t^{H}  \cr
& t^{H} & \Delta + 2\varepsilon + \frac{3J}{4} +U_M + V_{HL} + 2 V_{HM} + 2 V_{LM} },
\end{equation}
where the states above the matrix indicate the basis in which we write the matrix. The triplet $n_L=1$ block is
\begin{equation} \label{eq:HT1}
\hat{H}_T^{1} = \bordermatrix{ & \ket{^{3}MLCT^{1}} & \ket{^{3}LC^{1}} \cr
 & \Delta + \varepsilon + U_H + 2 V_{HL} + 2 V_{HM} + V_{LM} &  t^{H} \cr
 & t^{H} & \Delta + 2\varepsilon - \frac{J}{4} + U_M + V_{HL} + 2 V_{HM} + 2 V_{LM} \cr},
\end{equation} 
the singlet $n_L=2$ block is
\begin{equation} \label{eq:HS2}
\hat{H}_S^2= 
\bordermatrix{& \ket{^{1}MLCT^2} & \ket{^{1}MH^2} & \ket{^{1}LC^2} \cr
& 2 \Delta + U_H + U_L + 4 V_{HL}  &  \sqrt{2}t^{H} & 0 \cr
& \sqrt{2}t^{H} & 2 \Delta + \varepsilon +U_L + 2 V_{HL} + V_{HM} + 2 V_{LM} & \sqrt{2}t^{H} \cr
& 0 & \sqrt{2}t^{H} & 2\Delta + 2 \varepsilon + U_L + U_M + 4 V_{LM} },
\end{equation}
\end{widetext}
and there is only one state in the triplet $n_L=2$ block, the $\ket{^{3}MH^{2}}$ state, $\hat{H}_T^2 = (2\Delta + \varepsilon + U_L + 2 V_{HL} + V_{HM} + 2 V_{LM})$.
The coupling matrices are 
\[
\hat{T}^{0,1}_S \equiv \left( -\sqrt{2} t^L, 0 \right),
\]
\[
\hat{T}^{1,2}_S \equiv \left(\begin{array}{ccc}
	-\sqrt{2} t^L& 0& 0 \\
	0& t^L& 0
\end{array}\right),
\]
\[
\hat{T}^{0,2}_S = \left(0,0,0\right)
\]
amongst the singlet states, and for the triplets
\[
\hat{T}^{2,1}_T \equiv \left( 0 , t^L \right).
\]

\section{Approximate and exact solutions of the model Hamiltonian}
We will now investigate both the analytical solutions of Eq. \ref{eq:thehamiltonian11} in the limit of $t^{L} = 0, t^{H} \neq 0$ and the exact solutions numerically, i.e., we find the full configuration interaction solution, to gain some insight into the behavior of the model. 

A typical value of $\Delta$ is around 3 eV, much larger than the typical $t^L$ values of 0.1 eV (see Appendix). This means that the hybridization between states of different $n_L$ will be small compared to the hybridization between states of the same $n_L$, as is clear from Fig. \ref{fig:eigenstates}. As such, we expect the approximation that $t^{L}=0$ is close to the exact solution.
In the limit $t^L = 0$ all of the $\hat{T}^{n_L,n_L'}$ matrices are null, and hence $n_L$ is a good quantum number. 

The lowest excited states will come from the $n_L = 1$ subspace, provided $\Delta$ is large, as we expect it to be in complexes with OLED and OPV applications (see the Appendix). The eigenvalues of this subspace are
\begin{eqnarray} \label{eq:E1}
E^{(1)\pm}_{S,T} &=&  \frac{1}{2}\Bigg(2\Delta + 3\varepsilon^* +[3-4\mathbb{S}]\frac{J}{4} - 2 U_M + 4 U_H \nonumber \\
&& \qquad \, +\, 6 V_{HL} + 2 V_{HM}\Bigg) \nonumber \\
&& \pm \, \frac{1}{2} \Bigg[\Big(\varepsilon^* +[3-4\mathbb{S}]\frac{J}{4}\Big)^2 + 4(t^{H})^2 \Bigg]^{1/2},
\end{eqnarray}
where 
\begin{equation} \label{eq:epstar}
\varepsilon^* \equiv \varepsilon +U_M - U_H + V_{LM} - V_{HL}
\end{equation}
is the effective HOMO-metal energy gap relevant for the lowest excited states with $\mathbb{S} = 0 \, (\mathbb{S} = 1)$ for the eigenvalues of the singlet states (triplet states). $\varepsilon^*$, $J$ and $t^H$ are the key energy scales that determine the properties of the $n_L=1$ states. For states with purely $n_L=1$ character, varying $\varepsilon^*$ is equivalent to varying any one of $\varepsilon$, $U_M$, $U_H$, $V_{LM}$, or $V_{HL}$ as given by Eq. \ref{eq:epstar}. If we relax the approximation $t^L=0$ we need to deal with states with $n_L \neq 1$. If $t^L\neq0$, then in principle $\varepsilon^*$, $J$ and $t^H$ are no longer the only energy scales - one also needs to consider the individual effects of each of the direct Coulomb terms. For our typical parameter set, the effect of $t^L$ on the lowest excited states is quite small. This means that $\varepsilon^*$ is still the key parameter, as it is in the $t^L = 0$ case, so long as $\varepsilon^*$ does not vary too much (\textit{cf}. Supp. Info. Figs. S7-S17).\cite{suppinfo}

\begin{figure}
	\centering
		\includegraphics[width=0.9\columnwidth]{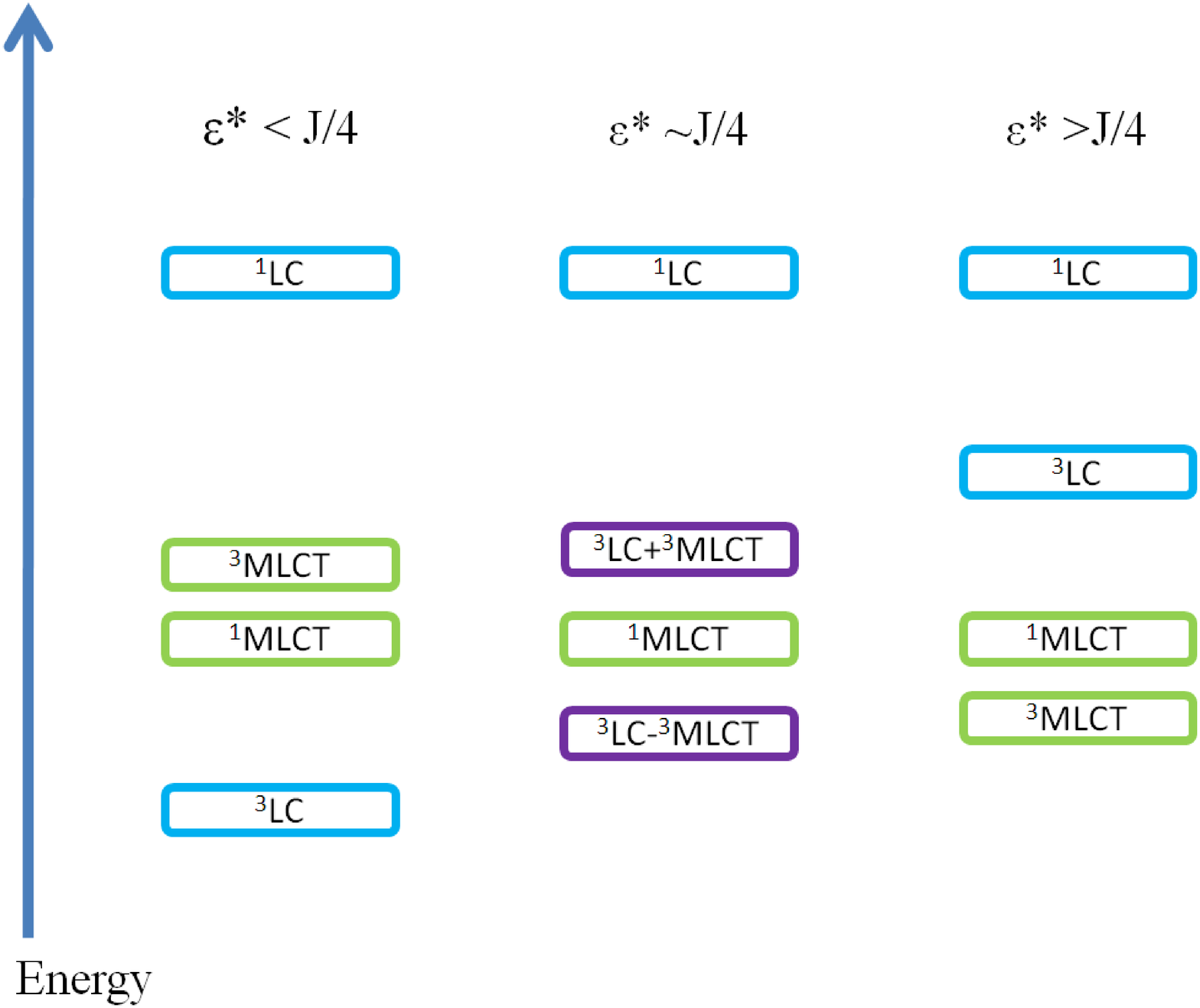}
	\caption{Schematic of the energy and character of the $n_L=1$ excited states in the regimes $\varepsilon^* < J/4$, $\varepsilon^* \sim J/4$ and $\varepsilon^* > J/4$, where $\varepsilon^* \equiv \varepsilon +U_M - U_H + V_{LM} -V_{HL}$. As $\varepsilon^*$ increases, the two triplet states approach the energy of the singlet MLCT state. At $\varepsilon^* = J/4$ there is an avoided crossing and the triplet states change character, with the lowest energy triplet state going from predominantly LC when $\varepsilon^* < J/4$, to predominantly MLCT when $\varepsilon^* > J/4$. This can be seen for some explicit parameter values in Fig. \ref{fig:onemetonelig_tripletmixing}.}
	\label{fig:loweststatecharacter}
\end{figure}

\begin{figure}[ht]
	\centering
		\includegraphics[width=0.9\columnwidth]{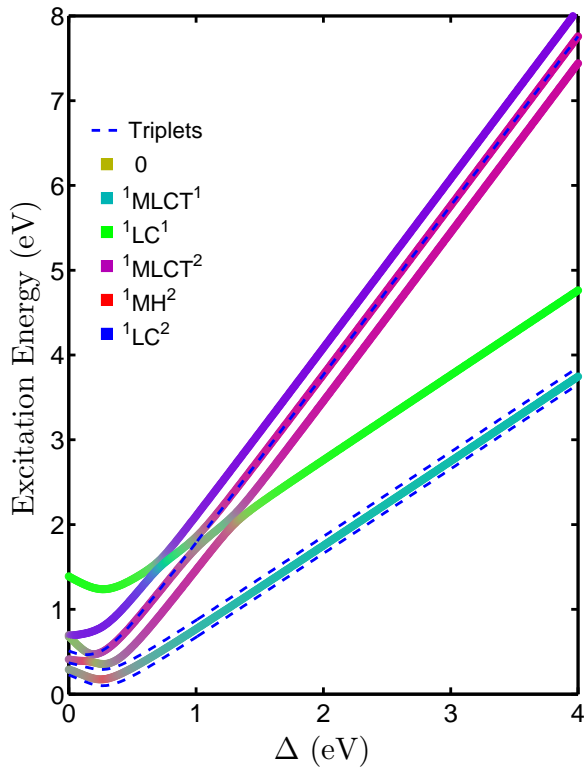}
		\caption{Singlet excitation energies as a function of the ligand HOMO-LUMO gap $\Delta$. The triplet excitation energies (dashed blue lines) are also shown for comparison. The colour of each curve shows how the character of the eigenstates changes with $\Delta$. Around the example value of $\Delta = 3$ eV there is no mixing of the lowest excited singlets, meaning they are nearly pure basis states. There are four pairwise avoided crossings and one avoided crossing involving three levels. These occur at the maximal hybridization conditions listed in \tab{HybridisationConditions}. Away from the avoided crossings, the eigenstates are predominantly the basis states for these values of the parameters. Note that all level crossings are avoided.  This plot is for our example parameter values (\textit{cf}. Fig. \ref{fig:eigenstates} and the appendix). Figs. S7, S9, S11, S12, S15, S16 and S17  in the Supp. Info. show that for the lowest excited states, this picture holds for a wide range of parameters around our typical parameter set.\cite{suppinfo}}
	\label{fig:onemetonelig_singletmixing}
\end{figure}

\begin{figure}[ht]
	\centering
		\includegraphics[width=0.9\columnwidth]{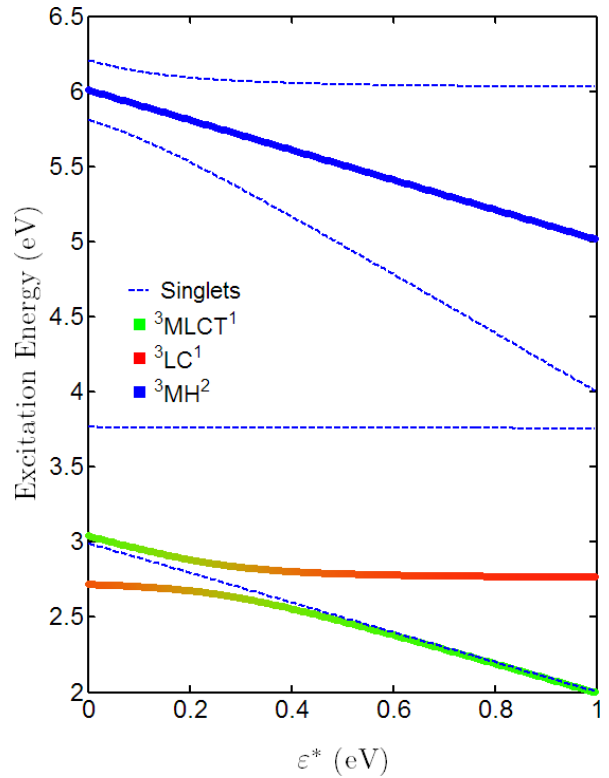}
		\caption{The triplet excitation energies as a function of the ligand HOMO-metal orbital gap $\varepsilon^*$. The singlet excitation energies (dashed blue lines) are also shown for comparison. The colour of each curve shows how the character of the eigenstates changes with $\varepsilon$. One can see the $^3$MLCT$^1$ - $^3$LC$^1$ hybridization point occurs at $\varepsilon^*=J/4 = 0.25$ eV as expected, while the $^3$MH$^2$ - $^3$LC$^1$ hybridization occurs at very large values of $\varepsilon^*$ for this parameter set. The location of the $^3$MLCT$^1$ - $^3$LC$^1$ maximum hybridization point at 0.25 eV puts it in the likely range of $\varepsilon$ values, meaning that in realistic circumstances the lowest triplet state will probably be a hybrid state. In contrast, avoided crossings in the lowest excited singlet state occur at $\varepsilon \gg 1$ eV, outside the reasonable range in these complexes. This plot is for our example parameter values (\textit{cf}. Fig. \ref{fig:eigenstates} and the appendix), varying $\varepsilon$ between 0 and 1 eV. Figs. S8, S10, S13 and S14 in the Supp. Info. show that this picture holds for a wide range of parameters around our typical parameter set, and reinforce the relevance of $\varepsilon^*$ as the key parameter for the $n_L=1$ excited states.\cite{suppinfo}}
	\label{fig:onemetonelig_tripletmixing}
\end{figure}

Fig. \ref{fig:loweststatecharacter} illustrates the character of the $n_L=1$ singlet and triplet states. If $(\varepsilon^* -J/4)/t^H$ is small then the LC and MLCT triplets will be strongly mixed (note that this does not require that $t^{H}$ is large compared to any of the parameters). On the other hand, in the absence of interactions, the level crossing would occur at $\varepsilon = 0$, the point at which the one electron $H^\dagger \ket{vac}$ and $M^\dagger \ket{vac}$ states are degenerate. Fig. \ref{fig:onemetonelig_singletmixing} and Fig. \ref{fig:onemetonelig_tripletmixing} show that the $n_L=1$ states are well separated from the other states for our typical parameter set (see Appendix for more details). It is important to stress that a large hybridization of the metal d and ligand $\pi$ orbitals is neither necessary nor sufficient to have a triplet excited state with mixed MLCT and LC character. This is different to what is discussed in Refs. \onlinecite{lo06} and \onlinecite{hay02}. It is well known\cite{ghosh06,cohen08} that density functional theory (DFT) tends to overestimate the delocalisation of electron density in excited states. As such the conclusions drawn from DFT with regards to changes in electron density, for example identifying transitions as MLCT or LC, may not be reliable.
\begin{table*}
	\centering
		\begin{tabular}{|c|c|c|}
		\hline
		States & Degeneracy Conditions & Coupling \cr
		\hline \hline
			$\ket{^{3}MLCT^{1}}$, $\ket{^{3}LC^{1}}$ & $\varepsilon^* = J/4$ & $t^{H}$ \cr
			$\ket{^{1}MLCT^{1}}$,  $\ket{^{1}LC^{1}}$ & $\varepsilon^* = -3 J/4$  & $t^{H}$ \cr
			$\ket{^{3}LC^{1}}$,  $\ket{^{3}MH^{2}}$ & $\Delta = \varepsilon^* - J/4 + U_H - U_L + V_{HM} - V_{LM}$ & $t^{L}$ \cr
			$\ket{^{1}LC^{1}}$,  $\ket{^{1}MH^{2}}$ & $\Delta = \varepsilon^* +3J/4 + U_H - U_L + V_{HM} - V_{LM}$ & $t^{L}$ \cr
			$\ket{0}$, $\ket{^{1}MLCT^{1}}$ & $\Delta = \varepsilon^* + U_H + 2 V_{HM} - V_{HL} - 2 V_{LM}$ & $\sqrt{2} t^{L}$ \cr
			$\ket{0}$, $\ket{^{1}MLCT^{2}}$ & $2 \Delta = 2 \varepsilon^* + 2 U_H - U_L - U_M + 4 V_{HM} - 2 V_{HL} - 2 V_{LM}   $ & $\sqrt{2} t^{L}$ \cr
			$\ket{^{1}MLCT^{1}}$,  $\ket{^{1}MLCT^{2}}$ & $\Delta = \varepsilon^* + U_H - U_L - U_M + 2 V_{HM} - V_{HL}  $ & $\sqrt{2} t^{L}$ \cr
			\hline
			$\ket{^{1}LC^{1}}$,  $\ket{^{1}MLCT^{2}}$ & $\Delta = 2 \varepsilon^* + 3J/4 + U_H - U_M - U_L - V_{HL} + 2 V_{HM} $ & indirect \cr
			$\ket{^{1}LC^{1}}$,  $\ket{^{1}LC^{2}}$ & $\Delta = 3J/4 - U_L + 2 V_{HM} + V_{HL} - 2 V_{LM}$ & indirect \cr
			$\ket{0}$,  $\ket{^{1}MH^{2}}$ & $2 \Delta = \varepsilon^* + 2 U_H - U_L + 3 V_{HM} - V_{HL} - 3 V_{LM}$ & indirect \cr	
			\hline		
			$H^\dagger\ket{vac}$ and $M^\dagger\ket{vac}$ & $\varepsilon = 0$ & $t^{H}$ \cr
			$L^\dagger\ket{vac}$ and $M^\dagger\ket{vac}$ & $\Delta = \varepsilon$ & $t^{L}$ \cr
			\hline
		\end{tabular}
	\caption{Level crossing points for many-body states and one electron states in the one metal one ligand model and the Hamiltonian element coupling them. Some of the hybridizations occur indirectly (\textit{i.e.} the states are not directly connected by a Hamiltonian element but can be at higher orders in perturbation theory). The first block of conditions are those between states with a Hamiltonian element directly coupling them (coupling at first order in perturbation theory). The second block are those between states which have no direct coupling in the Hamiltonian (higher orders in perturbation theory). The final block is of the hybridization conditions for the non-interacting model ($U_H = U_L = U_M = V_{HM} = V_{HL} = V_{LM} = J = 0$).} \label{tab:HybridisationConditions} 
\end{table*}

In the exact solution, neither $n_L$ nor $n_H$ are good quantum numbers. There is not much insight to be gained from the analytic solutions to this Hamiltonian, so we proceed by using some typical parameter values (discussed further in the Appendix) and investigating around these values numerically. All level crossings are avoided, as is clear from Figs. \ref{fig:onemetonelig_singletmixing} and \ref{fig:onemetonelig_tripletmixing}. These figures also show that the eigenstates are almost pure basis states everywhere except very near the avoided crossings (listed in \tab{HybridisationConditions}), since $t^H$ and $t^L$ are small compared to all the other Hamiltonian parameters. Hence, $\expect{n_L} \simeq 0,1$ or 2 for each of the singlet eigenstates, except near the avoided level crossings.

\tab{HybridisationConditions} lists all the (avoided) level crossings in our model, as well as the equivalent conditions for a non-interacting model. $J$ plays a key role in determining the level crossing for most of the states, so the singlet and triplet spectra are very different. By comparison, the level crossing points for the non-interacting states (one-electron states) only depend on the one electron site energies. Thus, singlets and triplets are degenerate in the non-interacting model. 
The differences between the mixing of the singlet states and that of the triplet states are apparent from Figs. \ref{fig:onemetonelig_singletmixing} and \ref{fig:onemetonelig_tripletmixing}. 

Fig. \ref{fig:onemetonelig_singletmixing} shows for our typical set of parameters how the eigenvalues change as a function of $\Delta$, and how the singlet eigenstates are dominated by one basis state except near avoided crossings. Note that for the typical parameter set no singlet avoided crossings occur for $\Delta>2$ eV, i.e. in the range of $\Delta$ values found in the ligands with applications in OLED and OPV devices (see Appendix). Fig. \ref{fig:onemetonelig_tripletmixing} shows the character of the triplet states as a function of $\varepsilon$, and how the triplet states can be strongly hybridized even away from avoided crossings. These results indicate the importance of the exchange interaction $J$ in determining the character and energy of the excited states. 
As is apparent from Figs. \ref{fig:onemetonelig_singletmixing} and \ref{fig:onemetonelig_tripletmixing}, for a wide range of reasonable parameter values the $S_1$ and $S_2$ states of our model are almost pure $\ket{^{1}MLCT^{1}}$ and $\ket{^{1}LC^{1}}$ basis states, respectively. As such, one can calculate the gap between these states to be $\varepsilon^* + 3J/4$. With an independent estimate of $J$ based on isolated ligand data (see Appendix A.1), one can estimate $\varepsilon^*$ from the energies of the $S_1$ and $S_2$ states found in absorption spectra, and compare it to the value found from the singlet-triplet gap (Eq. 17). This allows for a self-consistency check on the model, and an estimate of its accuracy.

\section{Effect of the MLCT character of the triplet excited state on its radiative properties}

Experimental data indicate that the emitting state is predominantly triplet. However, there has been considerable debate over the exact character of the state, \emph{i.e.}, whether it is LC, MLCT, or a hybrid.\cite{schwarz89,colombo93,lamansky01}
The degree of MLCT character in the triplet excited state has been discussed by Yersin \emph{et al.} as a key indicator of a compounds potential as an OLED emitter. \cite{yersinbook,yersin01,finkenzeller07} Here we examine how the MLCT character changes in the lowest singlet and triplet states, and how these changes in character might affect the radiative properties of the triplet states.

In an OLED device one wants to maximize the radiative decay rate compared to the non-radiative decay rate. It is not possible to predict how the non-radiative decay rate will vary from the current model, but we can examine how to increase the radiative rate via the increase of the transition dipole moment. For a triplet (phosphorescent) emitter, increasing the triplets transition dipole moment requires significant singlet-triplet mixing via spin-orbit coupling, which can only occur in the presence of the heavy transition metal core. \cite{kober82} The spin-orbit interaction couples electrons in atomic orbitals of different angular momentum, allowing spin flips between these orbitals and hence coupling singlets and triplets. In our model we only have one metal orbital (which is really some renormalized effective orbital we have identified as `metal'), so we insert a singlet-triplet coupling explicitly between states with an unpaired spin in the metal orbital to mimic the true effect of a spin-orbit interaction.

The triplet state $\ket{T_{m}}$ has a transition dipole moment $M_{T^m}$ to the ground state that is given (to first order in the spin orbit coupling Hamiltonian $\hat{H}_{SO}$) by
\begin{equation} \label{eq:socpert}
M_{T^m} = \sum_n \frac{\bra{T_{m}} \hat{H}_{SO}\ket{S_{n}}}{E_{T^m} - E_{S^n}} M_{S^n},
\end{equation}
where $E_{T^m}$ is the energy of the triplet state, $E_{S^n}$ is the energy of the singlet state $\ket{S_{n}}$ with transition dipole moment to the ground state $M_{S^n}$, and the sum runs over all singlet states (see page 271 of Ref. \onlinecite{hochstrasser} for more details). This expression is only valid if $\frac{\bra{T_{m}} \hat{H}_{SO}\ket{S_{n}}}{E_{T^m} - E_{S^n}} < 1$. If the singlet and triplet are nearly degenerate, then one needs to exactly solve the entire singlet-triplet Hamiltonian with explicit spin-orbit coupling.

\begin{figure}
	\centering
		\includegraphics[width=0.9\columnwidth]{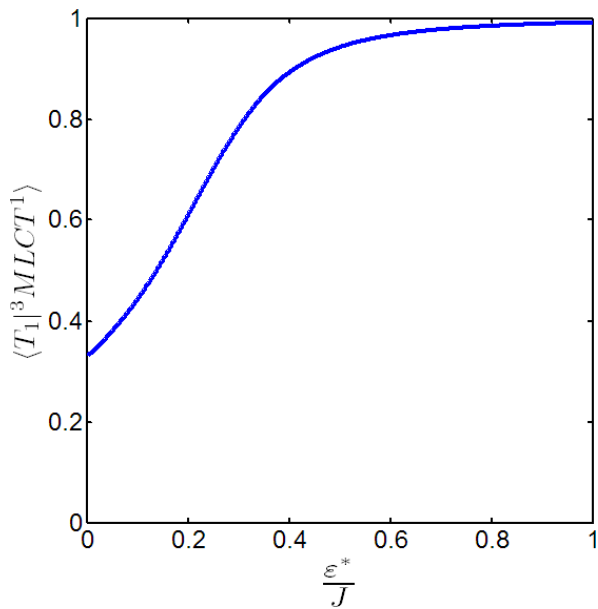}
	\caption{Degree of MLCT character in the lowest energy triplet state $\overlap{T_1}{{^3}MLCT^1}$ as a function of $\varepsilon^*/J$. As $\varepsilon^*/J$ increases, the MLCT state becomes favorable and begins to dominate the lowest triplet state. At the point $\varepsilon^*/J = 1/4$ the state is half MLCT, half LC ($\overlap{{^3}MLCT^1}{T_1}=\overlap{{^3}LC^1}{T_1}\simeq 1/\sqrt{2}$). Note that around our typical choice of parameters, varying $\varepsilon^*$ is equivalent to varying $\varepsilon$. To give the variation in $\varepsilon^*$ in this figure we use our typical parameter values (\textit{cf}. Fig. \ref{fig:eigenstates} and the Appendix), and vary $\varepsilon$ between 0 and 1 eV. }
	\label{fig:MLCTchar}
\end{figure}

Fig. \ref{fig:onemetonelig_tripletmixing} and Figs. S2-S6 of the Supp. Info.  show that the energy of the MLCT singlet lies between the two $n_L=1$ triplet states over a wide range of possible parameter values, while the LC singlet is higher in energy. These figures also show that when the lowest triplet state is predominantly MLCT (\textit{i.e.} $\varepsilon^* > J/4$), and is nearly degenerate with the singlet MLCT state. In this regime the perturbation expression above is no longer valid. For reasonable parameter values the MLCT singlet is the closest singlet (in energy) to the lowest triplet state (see Figs. \ref{fig:onemetonelig_singletmixing} and \ref{fig:onemetonelig_tripletmixing} and Figs. S7-S17 of the Supp. Info.).\cite{suppinfo} If we assume that the only contribution to the transition dipole moment of the lowest triplet state comes from the singlet which is nearly pure MLCT we can rewrite Eq. \ref{eq:socpert} as
\begin{eqnarray}
M_{T^1} &=& M_{^1 MLCT^1} \frac{\overlap{T_{1}}{{^3}MLCT^1}}{E_{T^1} - E_{^1 MLCT^1}} \nonumber \\
&& \times \bra{{^3}MLCT^1} \hat{H}_{SO} \ket{{^1}MLCT^1}. \label{eq:TDMT}
\end{eqnarray}
Fig. \ref{fig:MLCTchar} shows the amount of MLCT character in the lowest energy triplet state $\overlap{{^3}MLCT^1}{T_1}$ as a function of $\varepsilon^*/J$. 

We write our eigenstates as
\begin{equation}
\ket{S_i} \equiv c_{S_i}^{0} \ket{0} + c_{S_i}^{^1 MLCT^1} \ket{^1 MLCT^1} + ... ,
\end{equation}
so the $i^{\text{th}}$ singlet states transition dipole moment to the ground state $\ket{S_0}$ is
\begin{equation}
M_{S^i} \simeq e \sum_{\phi,\phi'} c_{S_i}^{\phi}c_{S_0}^{\phi'} \bra{\phi}\hat{r}\ket{\phi'}
\end{equation}
where $\ket{\phi},\ket{\phi'}$ are any of the basis states. We expect that due to the spatial separation between the metal and ligand orbitals, we can set $\bra{0}\hat{r}\ket{^1 MLCT^1} = 0$ and similarly for other terms involving separated excitations. The only terms we expect to remain are diagonal terms $\phi = \phi'$, and intraligand excitations \textit{i.e.} $\bra{0}\hat{r}\ket{^1 LC ^1}$.
In the range of our typical parameter values the lowest excited singlet state has almost pure $\ket{^1 MLCT^1}$ character, whereas the ground state is almost pure $\ket{0}$, so the dominant eigenstate coefficients will be $c_{S_0}^{0}$ and $c_{S_1}^{^1 MLCT^1}$. In the range of our typical parameter values, we find that $c_{S_1}^{0}c_{S_0}^{0} \simeq -c_{S_1}^{^1 MLCT^1}c_{S_0}^{^1 MLCT^1}$, so we can approximate the lowest singlet's transition dipole moment as 
\begin{equation} \label{eq:TDMS1}
M_{S^1} =  e c_{S_1}^{0}c_{S_0}^{0} \delta \vec{r},
\end{equation}
where $\delta \vec{r} \equiv \bra{^1 MLCT^1}\hat{r}\ket{^1 MLCT^1} - \bra{0}\hat{r}\ket{0}$.
Inserting Eq. \ref{eq:TDMS1} into Eq. \ref{eq:TDMT} we find
\begin{eqnarray}
M_{T^1} &=&  e c_{S_1}^{0}c_{S_0}^{0} \frac{\overlap{T_{1}}{{^3}MLCT^1}}{E_{T^1} - E_{^1 MLCT^1}} \nonumber \\
&& \times \delta \vec{r} \bra{{^3}MLCT^1} \hat{H}_{SO} \ket{{^1}MLCT^1}.
\end{eqnarray}
$e c_{S_1}^{0}c_{S_0}^{0} \frac{\overlap{T_{1}}{{^3}MLCT^1}}{E_{T^1} - E_{^1 MLCT^1}}$ will vary with our parameter values, while $ \delta \vec{r} \bra{{^3}MLCT^1} \hat{H}_{SO} \ket{{^1}MLCT^1}$ is a function only of our basis states and the strength of the spin-orbit coupling on the metal atom. 

The radiative decay rate of the triplet, found via the Einstein $A$ coefficient, is 
\begin{equation} \label{eq:lifetime}
\frac{1}{\tau_T} \equiv \frac{2 \omega^3}{3 \varepsilon_0 h c^3} \left|M_{T^1}\right|^2 
\end{equation}
where $\hbar \omega$ is the triplet-ground state energy gap and $\varepsilon_0$ is the permittivity of free space.\cite{hilborn02} This quadratic dependence on $M_{T^1}$ further amplifies the effects of small changes in $\varepsilon^*$. 

Apart from the Hamiltonian parameters, there is only one free parameter in the expression for the singlet lifetime (found from Eq. \ref{eq:TDMS1} via the Einstein $A$ coefficient), which is the `distance' of the transition, $\delta \vec{r}$. We solve our model with the typical parameter values and find that to reproduce a singlet radiative lifetime of $\sim$10 ns (see, for example, Ref. \onlinecite{gawelda07}) we must have $\delta \vec{r} = 20$ \AA. This distance is about twice the size of a Pd(thpy)$_2$ or Ir(ppy)$_3$ molecule. In other similar charge-transfer excitations, one often finds that the geometrical distance between the assumed `donor' and `acceptor' fragments is not well correlated with the dipole length.\cite{grisanti09}
In this case, the large value of $\delta \vec{r}$ may be due to our neglect of the $\bra{0}\hat{r}\ket{^1 LC ^1}$ term in calculation of the singlets transition dipole moment. The approximation that there is a large spatial separation between the metal and ligand orbitals may also be the source of the discrepancy, as the effective orbitals of our model may not be as localized as our labels suggest.

\begin{figure}
	\centering
		\includegraphics[width=0.9\columnwidth]{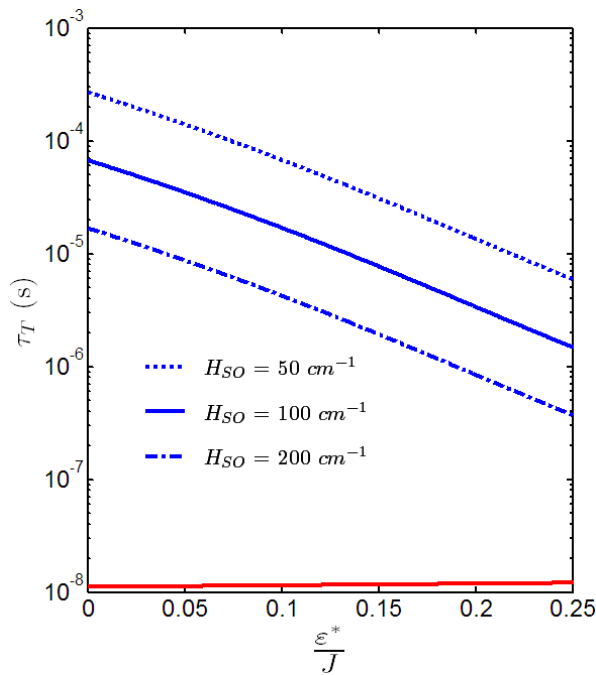}
	\caption{Triplet radiative lifetimes $\tau_T$ (upper three curves, blue) and singlet lifetimes (lowest curve, red) as a function of $\varepsilon^*/J$ for various values of $H_{SO}$ solved to first order in perturbation theory. We have chosen $ \delta \vec{r} = 20$ \AA $ $ to reproduce a singlet lifetime of order 10 ns. The triplet lifetime decreases rapidly as $\varepsilon^*/J$ increases, up until the point where the lowest singlet and triplet are nearly degenerate (\textit{i.e.} both MLCT) at which point the perturbative solution becomes invalid. As the strength of the spin-orbit coupling increases, the triplet lifetime rapidly decreases. We use our typical parameter values (\textit{cf}. Fig. \ref{fig:eigenstates} and the Appendix), varying $\varepsilon$ between 0 and 0.25 eV.}
	\label{fig:tripletlifetime}
\end{figure}

We choose a reasonable value for the strength of the spin orbit coupling, $\bra{{^3}MLCT^1} \hat{H}_{SO} \ket{{^1}MLCT^1}$ $\sim$ 100 cm$^{-1}$ = 0.012 eV (\emph{cf.} Ref. \onlinecite{bersuker}) and use the value $\delta \vec{r}$ = 20 \AA $ $ discussed above, and plot the calculated the lowest exited singlet and triplet states lifetimes as a function of $\varepsilon^*/J$ in Fig. \ref{fig:tripletlifetime}. 
Fig. \ref{fig:tripletlifetime} shows that the triplet lifetime $\tau_T$ changes rapidly as a function of $\varepsilon^*/J$ as long as $\varepsilon^*/J <  1/4$, \textit{i.e.} the triplet state is not within $\bra{{^3}MLCT^1} \hat{H}_{SO} \ket{{^1}MLCT^1}$ of the singlet energy (the condition for perturbation theory to be valid). Changing $\varepsilon^*/J$ from 0.1 to 0.25 decreases the triplet's radiative lifetime by an order of magnitude. For example with $\bra{{^3}MLCT^1} \hat{H}_{SO} \ket{{^1}MLCT^1} = 100$ cm$^{-1}$, the triplets radiative lifetime changes from 17 $\mu$s to 1.5 $\mu$s. These values of $\tau_T$ are comparable to those found experimentally in various organometallic complexes, for example $\sim$ 1 $\mu$s for Ir(ppy)$_3$ (Ref. \onlinecite{baldo99B}), or $\sim$ 100 $\mu$s for PtOEP (octaethyl-porphyrin platinum(II) (Ref. \onlinecite{baldo98}).

One can understand the apparently exponential change in $\tau_T$ as follows. We are explicitly in the regime where $\ket{T_1}$ is dominated by $\ket{^3 LC^1}$. If we treat the MLCT component of $\ket{T_1}$, $\overlap{T_{1}}{{^3}MLCT^1}$, perturbatively we find it contains a factor $(E_{T^1} - E_{^3 MLCT^1})^{-1} = (E_{T^1} - E_{^1 MLCT^1})^{-1}$. Thus the transition dipole moment of the triplet contains a factor $(E_{T^1} - E_{^1 MLCT^1})^{-2}$, so the lifetime varies with $(E_{T^1} - E_{^1 MLCT^1})^{4}$. Over the small regime we vary $\varepsilon^*$, $(E_{T^1} - E_{^1 MLCT^1})$ decreases linearly with $\varepsilon^*$, as seen in Fig. \ref{fig:onemetonelig_tripletmixing}. In this small range, a large power will be approximately linear on a semilogarithmic plot.

These results show that the excited state character and lifetime can be sensitively dependent on changes in Hamiltonian parameters. This implies that small changes to the chemistry of a complex, for example replacing a single hydrogen atom on a ligand molecule with a fluorine atom, could result in large changes in the molecules photophysical properties and, hence, the efficiency of optoelectronic devices made from these molecules. Observations of precisely this effect have been made in several systems, for example in a series of blue phosphorescent iridium complexes in Ref. \onlinecite{lo06}, and in a series of complexes based around N-heterocyclic cyano-substituted carbenes in Ref. \onlinecite{haneder08}. Ref. \onlinecite{haneder08} found that changing the ligand altered the singlet-triplet gap, and correlated this change in the gap with the changes in the radiative lifetime of the triplet state, finding the $\tau_T \propto (E_{T^1} - E_{^1 MLCT^1})^2$ relationship we predict in the regime where the spin-orbit coupling can be included perturbatively, \emph{i.e.}, for $\varepsilon^*/J < 1/4$.

In the regime $\varepsilon^*/J >  1/4$, where the lowest excited singlet and triplet are both of MLCT character, the dominant splitting between them comes from an exchange interaction between the metal orbital and the ligand orbitals, an interaction we have neglected in this model. The relative size of this inter-site exchange and the spin-orbital coupling element will determine the excited state properties in this regime.
Spin-orbit coupling may also have important effects on the non-radiative lifetime of the excited state, but these effects cannot be studied within the current model. One can expect that since non-radiative processes are thermally activated, small changes in energy barriers can cause exponentially larger changes in the rate.

\section{Charge injection}

An important problem to understand in organometallic complexes is the energetics of charge injection and extraction, and the influence of the character of the states involved.\cite{jacko10b} When electrons (holes) are injected into a bulk sample of our organometallic complex at an anode (cathode), a five (three) electron state is formed on a complex. When these oppositely ionized complexes are near each other they can return to charge neutral states (charge recombination) with one complex in its neutral ground state, and the other in a neutral excited state. It is the photoemission from this excited state which is desired in an OLED device. We would like to predict the relative population of each emitting excited state.

The typical lifetime of a singlet state is on the order of 10 ns, and for a triplet it is much longer (around 10 $\mu$s). It is believed that in complexes such as these a singlet-triplet intersystem crossing can occur in 50 fs.\cite{yoon06} This is more than 5 orders of magnitude greater than either excited state lifetime. As such, the population of the excited states has plenty of time to become a thermal population. This means we can predict the probability of finding either triplet or singlet excited states in the thermal population based on the energy difference between those states. 

The probability of the excited state being a triplet state at a temperature $T$ given by a Boltzmann distribution is 
\begin{equation} \label{eq:injtrip}
P_T = \frac{1}{1 + \frac{1}{3}e^{-(E^S - E^T)/(k_B T)} }
\end{equation}
whilst the probability of finding it in a singlet state is $P_S = 1-P_T$. 

\begin{figure}
	\centering
		\includegraphics[width=0.9\columnwidth]{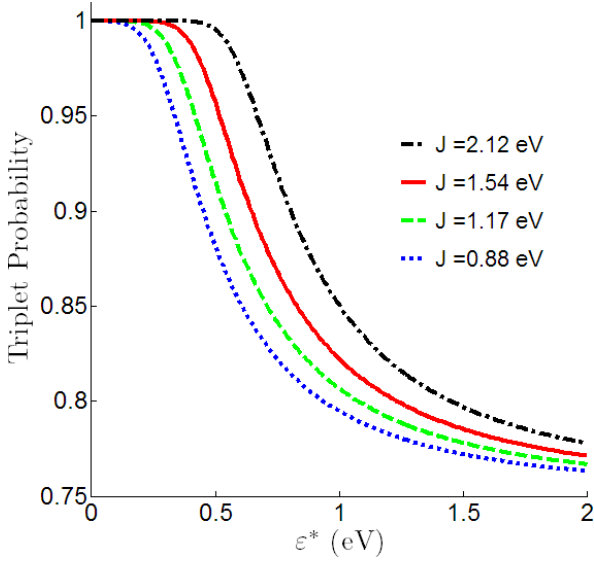}
	\caption{Probability of injected charges forming a triplet excited state versus $\varepsilon^*$ for various values of $J$. This probability is highly sensitive to $\varepsilon^*/J$. It is clear that increasing $\varepsilon^*$ suppresses triplet formation. The point at which the rapid decline from probability 1 begins is $\varepsilon^*/J =  1/4$, the point at which the triplet state gains large MLCT character, bringing its energy close to that of the lowest excited singlet state. The values of $J$ correspond to the ligands ppy ($J = 2.12$ eV), thpy ($J = 1.54$ eV), fluorene ($J = 1.17$ eV) and bzq ($J = 0.88$ eV), all at 300 K.  This plot was made with the typical parameter values (\textit{cf}. Fig. \ref{fig:eigenstates} and the appendix), varying $\varepsilon$ from 0 eV to 2 eV. }
	\label{fig:tripletprob_VSepsilonstar}
\end{figure}

Increasing $\varepsilon^*$ rapidly suppresses the triplet population, \emph{cf.} Fig. \ref{fig:tripletprob_VSepsilonstar}. As $\varepsilon^*$ increases, the probability of finding the excitation in the triplet state decreases, towards the limit of 75\%. The triplet probability depends on precisely what ligand and metal core one uses to construct the complex due to the probabilities sensitivity to $\varepsilon^*$. At $\varepsilon^* = 0.5$ eV on the curve with $J=$1.17 eV the probability of finding the excitation in the triplet state is above 90\%. The oft quoted proportion of 75\% triplets, 25\% singlets (see, for example, Ref. \onlinecite{sun06}) is only reached in the limits $\varepsilon^* / J \rightarrow \infty$ or $T\rightarrow \infty$.

In the limit $t^{L}=0$ we can find the lowest singlet (triplet) excited state by solving Eq. \ref{eq:HS1} (Eq. \ref{eq:HT1}). Thus we have the energy of the state with recombined charges forming an excited singlet
\begin{eqnarray} \label{eq:EcombS}
E^{S} &=& \frac{1}{2}\Bigg(2\Delta + 3\varepsilon^* +\frac{3J}{4} - 2 U_M + 4 U_H \nonumber \\
&& \quad  \, \,+\, 6 V_{HL} + 2 V_{HM}\Bigg) \nonumber \\
&& - \, \frac{1}{2}\sqrt{\left(\varepsilon^* +\frac{3J}{4} \right)^2 + 4(t^{H})^2} \nonumber
\end{eqnarray}
or an excited triplet
\begin{eqnarray} \label{eq:EcombT}
E^{T} &=& \frac{1}{2}\Bigg(2\Delta + 3\varepsilon^* -\frac{J}{4} - 2 U_M + 4 U_H \nonumber \\
&& \quad  \, \, +\, 6 V_{HL} + 2 V_{HM}\Bigg) \nonumber \\
&& - \, \frac{1}{2}\sqrt{\left(\varepsilon^* -\frac{J}{4} \right)^2 + 4(t^{H})^2}. \nonumber
\end{eqnarray}
The energy gap is
\begin{eqnarray}
E^S - E^T &=& \frac{1}{2} \Bigg[ J  +\sqrt{\left(\varepsilon^* -\frac{J}{4} \right)^2 + 4(t^{H})^2} \nonumber \\
 && \quad - \sqrt{\left(\varepsilon^* +\frac{3J}{4} \right)^2 + 4(t^{H})^2} \Bigg].
\end{eqnarray}

Thus we see again that small variations in parameter values can have large effects, in this case shifting the triplet probability exponentially as shown in Fig. \ref{fig:tripletprob_VSepsilonstar}.
As $\varepsilon^*$ increases, there are two competing effects on the photoluminescent efficiency. One is the suppression of triplet production down to 75\% seen in Fig. \ref{fig:tripletprob_VSepsilonstar}, and the other is the rapid decrease in triplet lifetime seen in Fig. \ref{fig:tripletlifetime}. The condition ($\varepsilon^*\gg J$) necessary for a large triplet transition dipole moment also suppresses the probability of the formation of a triplet excited state.

\section{Conclusions}

We have investigated an effective model Hamiltonian for organometallic complexes in electronic devices. We have seen that while the lowest singlet state is typically a nearly pure MLCT state, the lowest triplet states character varies, changing from LC to MLCT with a highly hybridized region between. This variation in triplet character is strongly dependent on the ratio $\varepsilon^*/J$. Importantly, $\varepsilon^*$ is  purely a property of the complex and will depend sensitively on the ligand chemistry. The strong LC-MLCT mixing in the lowest triplet state means that a small shift in parameter values can cause large changes in the properties of the state (changing $\varepsilon^*$ by a factor of 2 changes the triplet lifetime by almost an order of magnitude). This sensitive dependence provides an explanation for the large observed changes in the photophysical properties of organometallic complexes caused by small changes in the ligands (such as changing a single substituent atom on the ligand). As well as having a direct effect on the lifetime, the change in excited state energy which accompanies the change in hybridization causes a shift in the probability of finding the excitation in the triplet state. As $\varepsilon^*$ increases, the triplet decay rate increases by orders of magnitude while the triplet probability decreases by at most 33\%.

\begin{acknowledgments}
We are grateful to Arthur Smith for critically reading this manuscript, and to Paul Burn, Arthur Smith, Paul Shaw, Lawrence Lo and Seth Olsen for helpful discussions.
B. J. P. was the recipient of an Australian Research Council (ARC) Queen Elizabeth II Fellowship (project no. DP0878523). R. H. M. was the recipient of an ARC Australian Professorial Fellowship (project no. DP0877875).
\end{acknowledgments}

\appendix

\section{Estimated parameter values} \label{App_param}
We would like to understand the values of, and relationships between, the Hamiltonian parameters ($J$, $\Delta$, $t^{H}$, $t^{L}$, $U_H$, $U_L$, $U_M$, $V_{HL}$, $V_{HM}$, $V_{LM}$).

We begin with a two site extended Hubbard model, modeling the ligand as two sites (for example the two phenyl rings in a biphenyl ligand) occupied by two electrons
\begin{widetext}
\begin{equation}
\hat{H} = -t_\ell \sum_\sigma \left( c_{1\sigma}^\dagger c_{2\sigma} + c_{2\sigma}^\dagger c_{1\sigma} \right) + U_\ell \left( n_{1\uparrow} n_{1\downarrow} + n_{2\uparrow}n_{2\downarrow} \right) + V_\ell \sum_{\sigma, \, \sigma'} n_{1\sigma} n_{2\sigma'}
\end{equation}
with $n_{i\sigma} \equiv c^\dagger_{i\sigma} c_{i\sigma}$. By neglecting an exchange interaction between these localized orbitals, we are making the CNDO (complete neglect of differential overlap) approximation in this small basis set.\cite{pople65}
By transforming to a basis of delocalised states $H^\dagger_{\sigma} \equiv \frac{c_{1\sigma}^\dagger + c_{2\sigma}^\dagger}{\sqrt{2}}$ and $L^\dagger_{\sigma} \equiv \frac{c_{1\sigma}^\dagger - c_{2\sigma}^\dagger}{\sqrt{2}}$, one finds that \cite{michl87}
\begin{eqnarray}\label{eq:Htwositedelocal}
\hat{H} &=& -t_\ell (n_H - n_L)  - J \vec{S}_H \cdot \vec{S}_L -\frac{J}{2}\left( H^\dagger_{\uparrow}H^\dagger_{\downarrow}L_{\uparrow}L_{\downarrow} + L^\dagger_{\uparrow}L^\dagger_{\downarrow}H_{\uparrow}H_{\downarrow} \right)\nonumber \\
 &&+ \, U_{H}(n_{H\uparrow}n_{H\downarrow} + n_{L\uparrow}n_{L\downarrow}) + V_{HL} \left( n_{H\uparrow}n_{L\downarrow}+ n_{H\downarrow}n_{L\uparrow} + n_{H\uparrow} n_{L\uparrow} + n_{H\downarrow} n_{L\downarrow}\right),
 \end{eqnarray}
\end{widetext}
where
\begin{eqnarray}
J &=& U_\ell - V_\ell, \label{eq:hubbardgetJ} \\
U_H = U_L &=& \frac{U_\ell + V_\ell}{2}, \\
V_{HL} &=& \frac{U_\ell + 3 V_\ell}{4}.
\end{eqnarray}

\subsection{Intraligand Exchange: $J \simeq 1$ eV:} 

The first excited singlet state energy of the two-site Hubbard model for the ligand is $E_S = V_{HL}+\frac{3J}{4}$, and the first excited triplet state energy is $E_T = V_{HL}-\frac{J}{4}$. Thus we can easily estimate $J$ as
\begin{equation} \label{eq:fitJ}
J = E_S - E_T.
\end{equation}
\begin{table}
	\centering
		\begin{tabular}{|c|c|c|c|c|c|}
		\hline
		Ligand	& Ref. & $S_{1}$ (eV) &	$T_{1}$ (eV) &	$J$ (eV) &	$\Delta$ (eV) \\ \hline
		thpy & \onlinecite{maestri85} & 4.08 &	2.54 &	1.54 &	3.31 \\
		ppy	 & \onlinecite{maestri85} & 4.99 &	2.87 &	2.12 &	3.93 \\
		bzq	 & \onlinecite{maestri85} & 3.57 &	2.69 &	0.88 &	3.13\\ 
		biphenyl  & page 108 of \onlinecite{handbookphoto} & 4.33	& 2.84	& 1.49 & 3.59 \\
		carbazole & page 111 of \onlinecite{handbookphoto} & 3.60 &	3.05 &	0.55 &	3.33 \\
		fluorene  & page 118 of \onlinecite{handbookphoto} & 4.11 &	2.94 &	1.17 &	3.53 \\
		\hline \end{tabular}
		\caption{Data for several common ligands. Experimental energies for the lowest visible singlet and triplet states, the calculated spin exchange $J$ using Eq. \ref{eq:fitJ} and HOMO - LUMO gap $\Delta$ using Eq. \ref{eq:fitDelta}.} \label{tab:ligdata}
\end{table}
\tab{ligdata} shows the value of $J$ estimated in this way for several common ligands. This data shows that using a value of $J \simeq 1$ eV is realistic for investigating our model. While the possible values of $J$ vary widely, the effects of this variation in the lowest excited states are identical to variations in $\varepsilon^*$. 
Rusanova \emph{et al.} found exchange interactions of the same magnitude (around 1 eV), having performed a semi-empirical INDO/S analysis on a series of ruthenium complexes.\cite{rusanova06} They also assert that the degree of singlet-triplet splitting is a measure of $\pi$ backbonding. This feature is naturally reproduced in our model as the only way to have singlet-triplet splitting in our model is via the $LC^1$ states, the only states directly split by $J$.

Figs. S9 and S10 of the Supp. Info. show that our specific choice of $J$ has no effect on the qualitative conclusions drawn in this paper.\cite{suppinfo}

\subsection{$H$-$L$ Splitting: $\Delta \simeq 3$ eV:} 

The energy difference between $H^\dagger_\uparrow H^\dagger_\downarrow \ket{0}$ and $L^\dagger_\uparrow L^\dagger_\downarrow \ket{0}$ is $2t_\ell$, and they are coupled by $J/2$. Thus the eigenstates will be separated by an energy $\sqrt{4 t_\ell^2 + J^2/4}$.  \tab{ligdata} shows that $J/2$ is around 0.5 eV, while the excited singlet to ground state gap is around 4 eV. We know the gap $\sqrt{4 t_\ell^2 + J^2/4}$ will be greater than the singlet ground state gap of $\sim 4$ eV. Since we know $J/2$ is an order of magnitude smaller than this energy, we make the approximation that the states $H^\dagger_\uparrow H^\dagger_\downarrow \ket{0}$ and $L^\dagger_\uparrow L^\dagger_\downarrow \ket{0}$ are eigenstates (implying that $\Delta = t_\ell$). We find the ground state is $H^\dagger_\uparrow H^\dagger_\downarrow \ket{0}$ with energy $-\Delta + U_H$. The ground state-singlet gap is $\Delta_S = \frac{J}{2}+\Delta$, and the ground state-triplet gap is $\Delta_T = -\frac{J}{2}+\Delta$ (using $U_H - V_{HL} = J/4$). Thus we find
\begin{equation} \label{eq:fitDelta}
\Delta = \frac{\Delta_S+\Delta_T}{2}.
\end{equation}
We apply this to spectral data from isolated ligands to estimate the value of $\Delta$ in \tab{ligdata}, finding that $\Delta \simeq 3$ eV is a realistic value to use in the investigations of the properties of the model Eq. \ref{eq:thehamiltonian11}.

Figs. S7, S9, S11, S12, S15, S16 and S17 of the Supp. Info. show that our specific choice of $\Delta$ within the range of possible values in \tab{ligdata} has no effect on the lowest excited states, and therefore no effect on the conclusions drawn here.\cite{suppinfo}

\subsection{Direct Coulomb Interactions on the Ligand: $U_H \simeq V_{HL} \simeq 3$ eV:} 

There is an empirical relationship between the $U_\ell$ on the localized sites and the $V_\ell$ between the localized sites, in terms of the inter-site spacing $R$,
\begin{equation} \label{eq:applocalcoloumb}
V_\ell^{-1} = R + U_\ell^{-1}
\end{equation}
in atomic units (page 20 of Ref. \onlinecite{fulde}).
\begin{figure}
	\centering
		\includegraphics[width=0.9 \columnwidth]{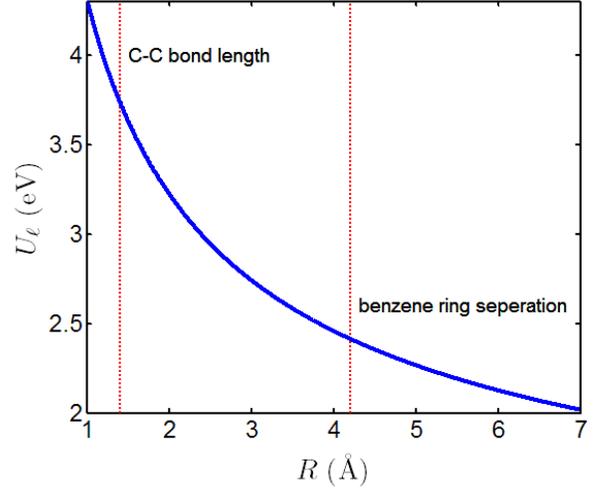}
	\caption{Predicted $U_\ell$ as a function of $R$ with the constraint that we match our typical value of $J = U_\ell-V_\ell \simeq 1$ eV. The dotted vertical red lines are at the carbon-carbon bond length in benzene 1.4 \AA = 2.65 $a_B$ and the distance between benzene ring centers in biphenyl, 3.2 \AA.}
	\label{fig:fuldestyle_RJpredictU}
\end{figure}
If we substitute Eq. \ref{eq:hubbardgetJ} we find that
\begin{equation}
U_\ell = \frac{J}{2} + \sqrt{\frac{J^2}{4}+\frac{J}{R}}.
\end{equation}
Fig. \ref{fig:fuldestyle_RJpredictU} shows that $U_\ell$ should be somewhere between 2.5 eV and 3.75 eV, given the possible range of $R$ and a typical value of $J = 1$ eV. It is worth noting here that a factor of three change in bond length is only a factor of 1.5 in the magnitude of $U$.

For $J=1$ eV we have
\begin{equation}
U_{H} =U_\ell - \frac{1}{2}\text{eV},\quad V_{HL} =U_\ell - \frac{3}{4}\text{eV},
\end{equation}
hence
\begin{equation}
U_H  = V_{HL}+\frac{1}{4} \text{eV}.
\end{equation}
We expect $U_H$ to be in the range 3.25 eV to 2 eV (based on the above range for $U_\ell$).
$U_H$ will be much larger than 0.25 eV, and hence we can approximate $U_H \simeq V_{HL}$. This analysis makes it seem reasonable to choose $U_H = V_{HL} = 3$ eV for our typical parameter set used to investigate the model. Ref. \onlinecite{rusanova06} evaluates direct Coulomb integrals for a series of Ru complexes, finding values $ \sim 4.5$ eV (calculated via semi-empirical INDO/S computations), similar to the values found from the above discussion. As discussed in Ref. \onlinecite{scriven09}, this kind of calculation is at best a reasonable upper bound for the value of the parameters in an effective low energy Hamiltonian.

Figs. S7 and S8 of the Supp. Info. show that relaxing the assumption that $U_H  = V_{HL}$ does not cause any qualitative changes in the solutions of the model.\cite{suppinfo}

\subsection{Direct Coulomb Interactions Involving the Metal Site: $U_M \simeq U_H, V_{HL} \simeq V_{HM}$:} 

Since the HOMO and LUMO are in the same location and have the same on-site Coulomb repulsion, we find that the intersite Coulomb repulsion between the ligand orbitals and the metal will be equal,
\begin{equation}
V_{HM}^{-1} = V_{LM}^{-1} = R_{HM}+\frac{2}{U_H + U_M}.
\end{equation}
We expect that $R_{12} \simeq R_{HM}$ (where $R_{12}$ is the distance between the two sites of our model of the ligand, and $R_{HM}$ is the distance between the ligand and the metal). As long as $U_M \sim U_H$, we will have $V_{HM} = V_{LM} \simeq V_{HL} = U_H$.
This reduces the six parameters for the direct Coloumb integrals ($U_H$, $U_L$, $U_M$, $V_{HL}$, $V_{HM}$, $V_{LM}$) to just two ($U_H$ and $U_M$). 

If we were to assume that $U_H \simeq U_M $ then each four-electron basis state gains an energy $6 U_H$ relative to the case with no Coulomb interactions. Thus it is clear that in this approximation the direct Coulomb interactions have no qualitative effect on the solutions to this Hamiltonian in the four electron subspace.

For our typical parameter values, the $n_L=1$ states are well separated from the $n_L \neq 1$ states. This means that while we are investigating the lowest excited states (the $n_L=1$ states) varying $\varepsilon^*$ captures all the same physics as varying $\varepsilon$, $U_M$, $U_H$, $V_{LM}$ and $V_{HL}$ individually. As such, it is convenient to choose $U_M = U_H$ and $V_{HL} = V_{HM}$ and then investigate the effects of changing $\varepsilon^*$.

Figs. S1-S4, S6-S8, S11-S17 of the Supp. Info. show that varying $\varepsilon$ and $U_M$ (equivalent to varying $V_{LM}$) cause no qualitative changes to the solutions of the model in reasonable parameter ranges.\cite{suppinfo} We must increase $U_M$ more than 3 eV above the typical value before there are any qualitative changes to the lowest excited states which would alter the conclusions drawn here (see Figs. S3 and S4).\cite{suppinfo}

\begin{table}[bh]
	\centering
		\begin{tabular}{|c|c|c|c|}\hline
		Pt & thpy \cite{breu97} & ppy \cite{chassot84} & bzq \cite{jolliet96}  \\ \hline
		$t^{H}$  &  0.09 eV & 0.08 eV & 0.08 eV \\ 
		$t^{L}$ &     0.11 eV & 0.11 eV & 0.06 eV  \\ \hline
		\end{tabular}
	\caption{Values of $t^{H}$ for various platinum complexes, estimated with a semi-empirical parameterization and using HOMOs and LUMOs from H\"uckel model calculations of isolated ligands using the parameterization on page 284 of Ref. \onlinecite{loweandpeterson}. The ligand - metal bond lengths for the complexes come from crystallographic data contained in the references in the table.}
	\label{tab:tvalues}
\end{table}

\subsection{Hopping Intergrals: $t^{H} \simeq t^{L} \simeq 0.1$ eV:} Using the standard semi-empirical parameterization (for example, page 551 of Ref. \onlinecite{harrison89}), along with H\"uckel HOMO and LUMO orbitals of an isolated ligand and experimental carbon-metal and nitrogen-metal bond lengths, we estimate the values of $t^{H}$ and $t^{L}$ for various ligands given in \tab{tvalues}. Note that the variation in the $t$ values is almost completely due to the differences between the H\"uckel orbitals (the variation due to the different experimental bond lengths is $\sim1$\%).

\subsection{HOMO-metal splitting, $\varepsilon$, is a property of the complex:} $\varepsilon$ is a property only of the whole complex which is difficult to predict \emph{a priori}. In Fig. \ref{fig:onemetonelig_singletmixing} we choose a value of $\varepsilon = 0.25$ eV for the sake of concreteness. Figs. S6, S15, S16 and S17 in the Supp. Info. show that this choice has no effect on our conclusions regarding the lowest excited states.\cite{suppinfo}

Figs. S1-S4, S6-S8, S11-S17 of the Supp. Info. show that varying $\varepsilon$ and $U_M$ and hence $\varepsilon^*$ cause no qualitative differences in the solutions of the model, as discussed above in the section on direct Coulomb integrals involving the metal site.\cite{suppinfo}

\end{document}


\title{Supplementary Information for `Sensitivity of the photo-physical properties of organometallic complexes to small chemical changes' : Robustness of parameter choice}
\author{A. C. Jacko}
\affiliation{Centre for Organic Photonics and Electronics, School of Mathematics and Physics, The University of Queensland}
\author{B. J. Powell}
\affiliation{Centre for Organic Photonics and Electronics, School of Mathematics and Physics, The University of Queensland}
\author{Ross H. McKenzie}
\affiliation{Centre for Organic Photonics and Electronics, School of Mathematics and Physics, The University of Queensland}
\date{\today}

\maketitle

One might be concerned that the conclusions we drawn from numerics at `typical parameter values' are not robust to variations in those parameters. Here we show that nothing in the lowest excited states qualitatively changes when we vary the parameters in a reasonable range. The typical parameter values around which we vary are $J = 1$ eV, $\Delta = 3$ eV,  $\varepsilon =0.25$ eV,  $t^H = 0.1$ eV,  $t^L = 0.1$ eV,  $U_M=U_H=U_L=U= 3$ eV,  $V_{HL} = V_{HM} = V_{LM} = V= 3$ eV.

Figs. S1-S4 show that the assumption $U_M = U_H$ does not qualitatively change the lowest excited singlet state until $U_M > 5$ eV. 
Fig. S5 shows that the approximation $t^{L} = 0$ is a valid approximation.
Fig. S6 shows that increasing the value of $\varepsilon$ has no significant effect on the lowest excited singlet state.
Figs. S7 and S8 show that the approximation $V=U$ does not have any strong effect on the character of the eigenstates. 
Figs. S9 and S10 show that the value of $J$ chosen does not have any strong effect on the character of the singlet eigenstates, but does change the value of $\varepsilon$ at which the lowest triplet state changes character. 
Figs. S11 and S12 show that the approximation $U_M=U_H$ does not have any strong effect on the character of the lowest singlet state. 
On the other hand, Figs. S13 and S14 show that in the triplet states, increasing $U_M-U_H$ shifts the $\ket{^3 MLCT^1}$-$\ket{^3 LC^1}$ degeneracy condition to lower values of $\varepsilon$ (equivalent to decreasing $J$).) 
Figs. S15 and S16 show that the exact value of $\varepsilon$ chosen has no effect on our conclusions regarding the lowest excited states. 
Fig. S17 shows that in the lowest excited states varying $U_M$ is equivalent to varying $\varepsilon$, as expected since $\varepsilon^*$ is the controlling parameter of the $n_L$ states.

\begin{figure}
	\centering
		\includegraphics[width=0.9\columnwidth]{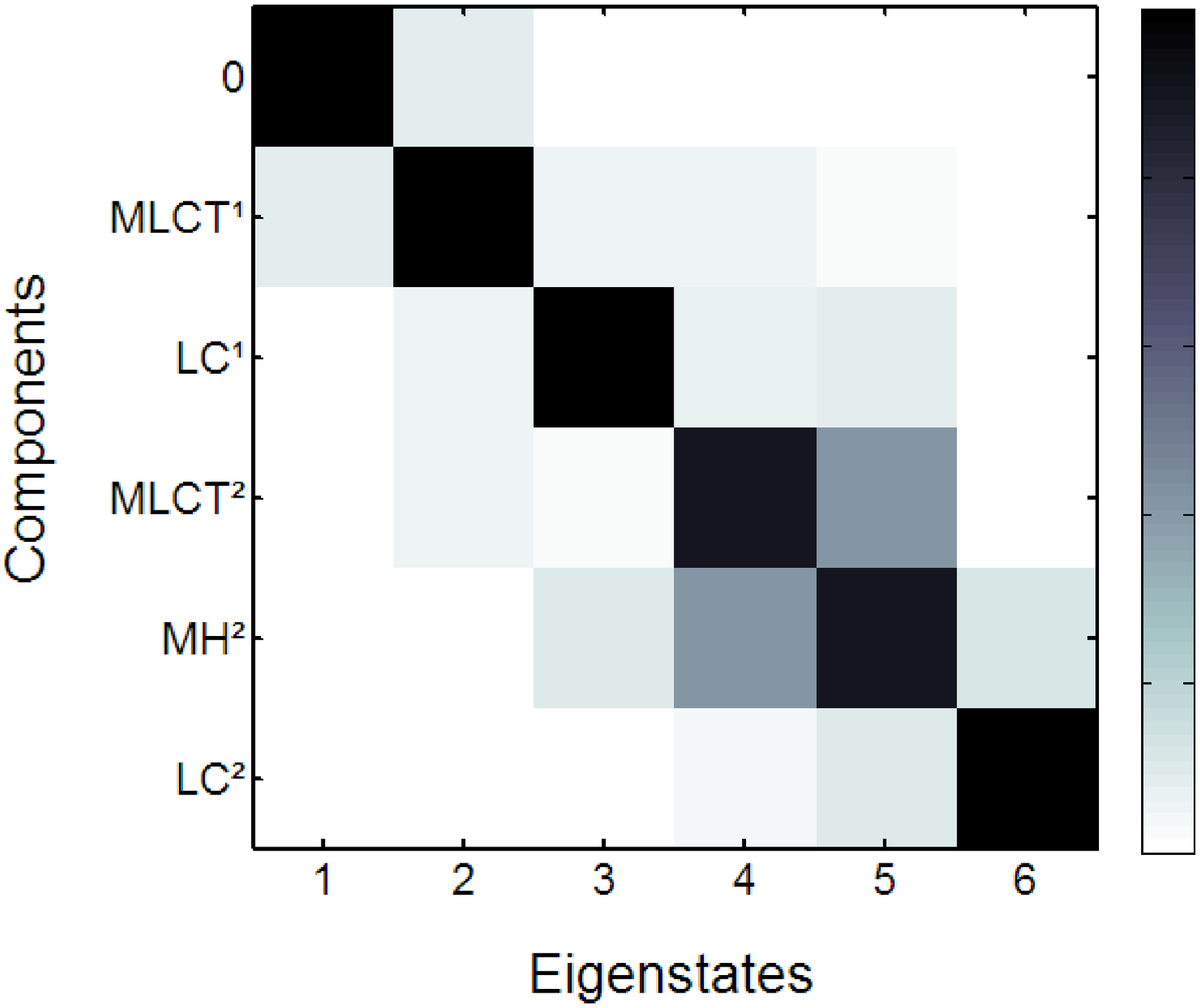}
		\caption{This figure shows the singlet eigenstates with $U_M = 4$ eV, for comparison with Fig. 2, showing that there is very little change in the lowest singlet states.}
  \label{fig:robust_eigenstates_UM_4}
\end{figure}

\begin{figure}
\centering
  \includegraphics[width=0.9\columnwidth]{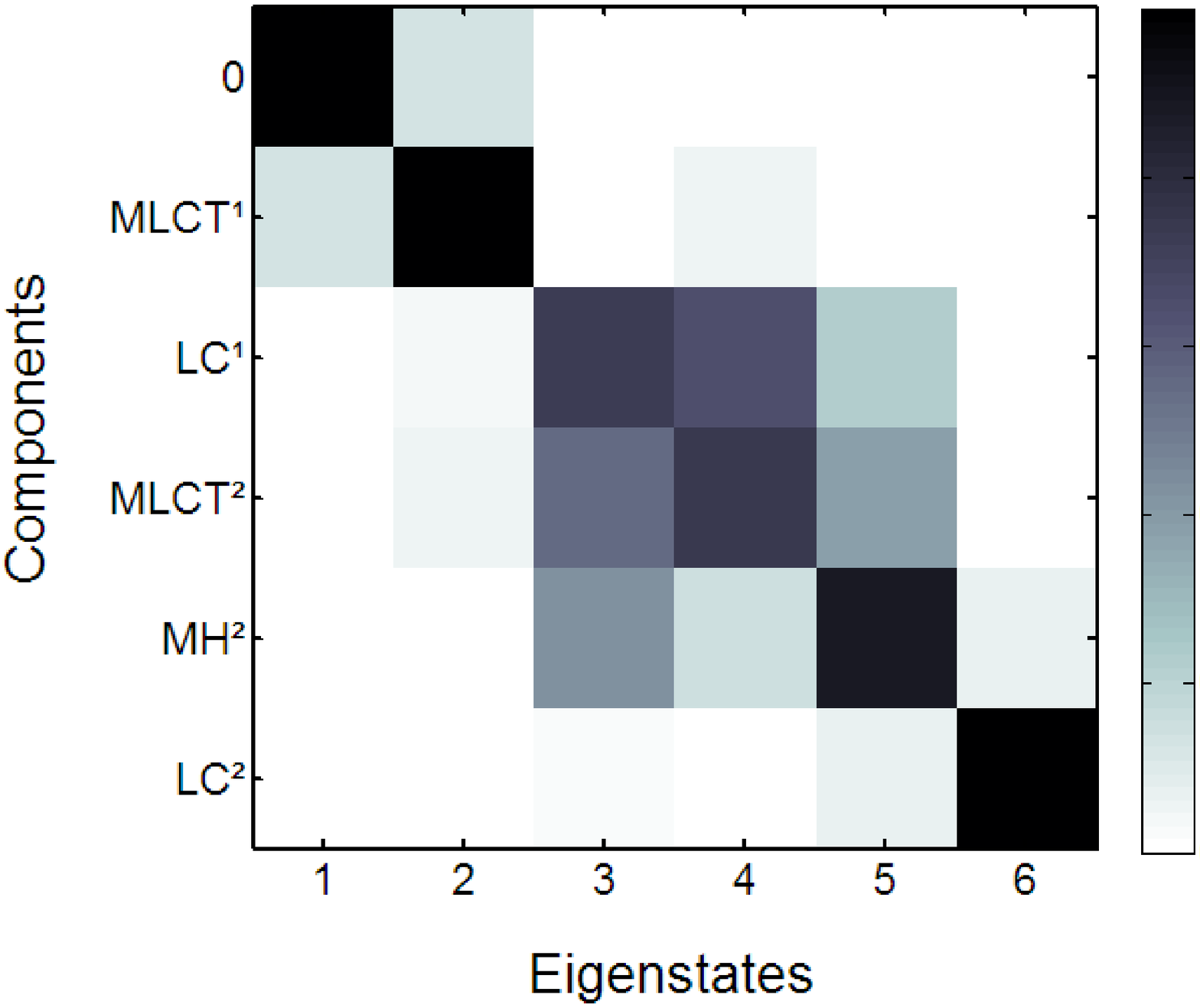}
  \caption{This figure shows the singlet eigenstates with $U_M = 4.7$ eV, for comparison with Fig. 2, showing that although the $\ket{^1 LC^1}$ and $\ket{^1 MLCT^2}$ states are mixing strongly, the ground state remains $\ket{0}$ and the first excited singlet remains predominantly $\ket{^1 MLCT^1}$. This shows that varying $U_M$ (and hence $\varepsilon^*$) by 1.7 eV has no qualitative effect on the lowest two singlet states.}
  \label{fig:robust_eigenstates_UM_4point7}
\end{figure}

\begin{figure}
\centering
		\includegraphics[width=0.9\columnwidth]{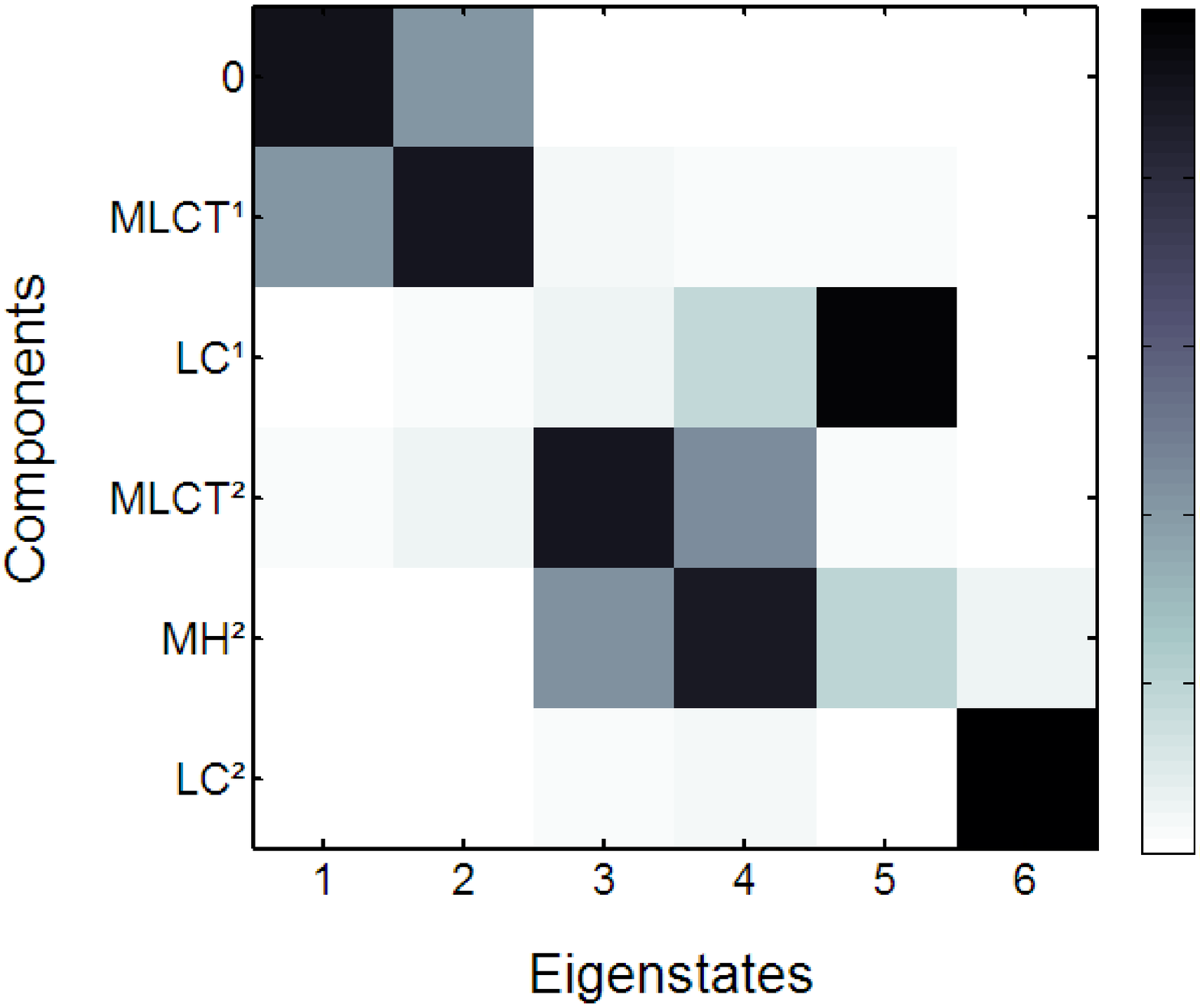}
		\caption{This figure shows the singlet eigenstates with $U_M = 5.5$ eV, for comparison with Fig. 2, showing that the $\ket{0}$ and $\ket{^1 MLCT^1}$ states are mixing more strongly, and that the $\ket{^1 LC^1}$, $\ket{^1 MLCT^2}$ and $\ket{^1 MH^2}$ states have changed order. Even so, the lowest singlet states have not qualitatively changed.}
  \label{fig:robust_eigenstates_UM_5point5}
\end{figure}

\begin{figure}
\centering
		\includegraphics[width=0.9\columnwidth]{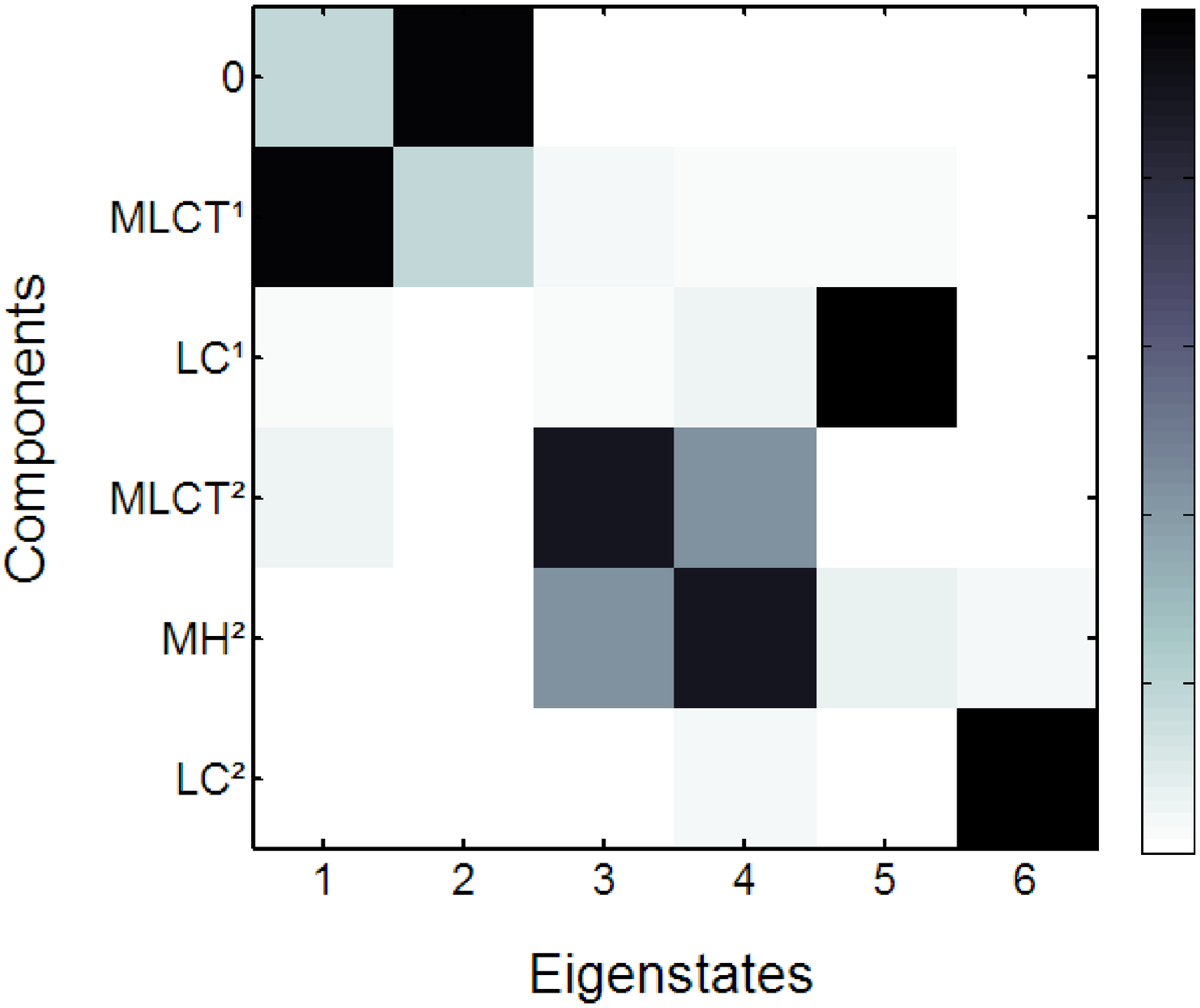}
		\caption{This figure shows the singlet eigenstates with $U_M = 6.5$ eV, for comparison with Fig. 2, showing that the $\ket{0}$ and $\ket{^1 MLCT^1}$ states have changed order, with the ground state now predominantly $\ket{^1 MLCT^1}$. This large qualitative change occurs 3.5 eV above our example value of $U_M$, so we do not expect this ordering to be physically relevant.}
  \label{fig:robust_eigenstates_UM_6point5}
\end{figure}

\begin{figure}
\centering
		\includegraphics[width=0.9\columnwidth]{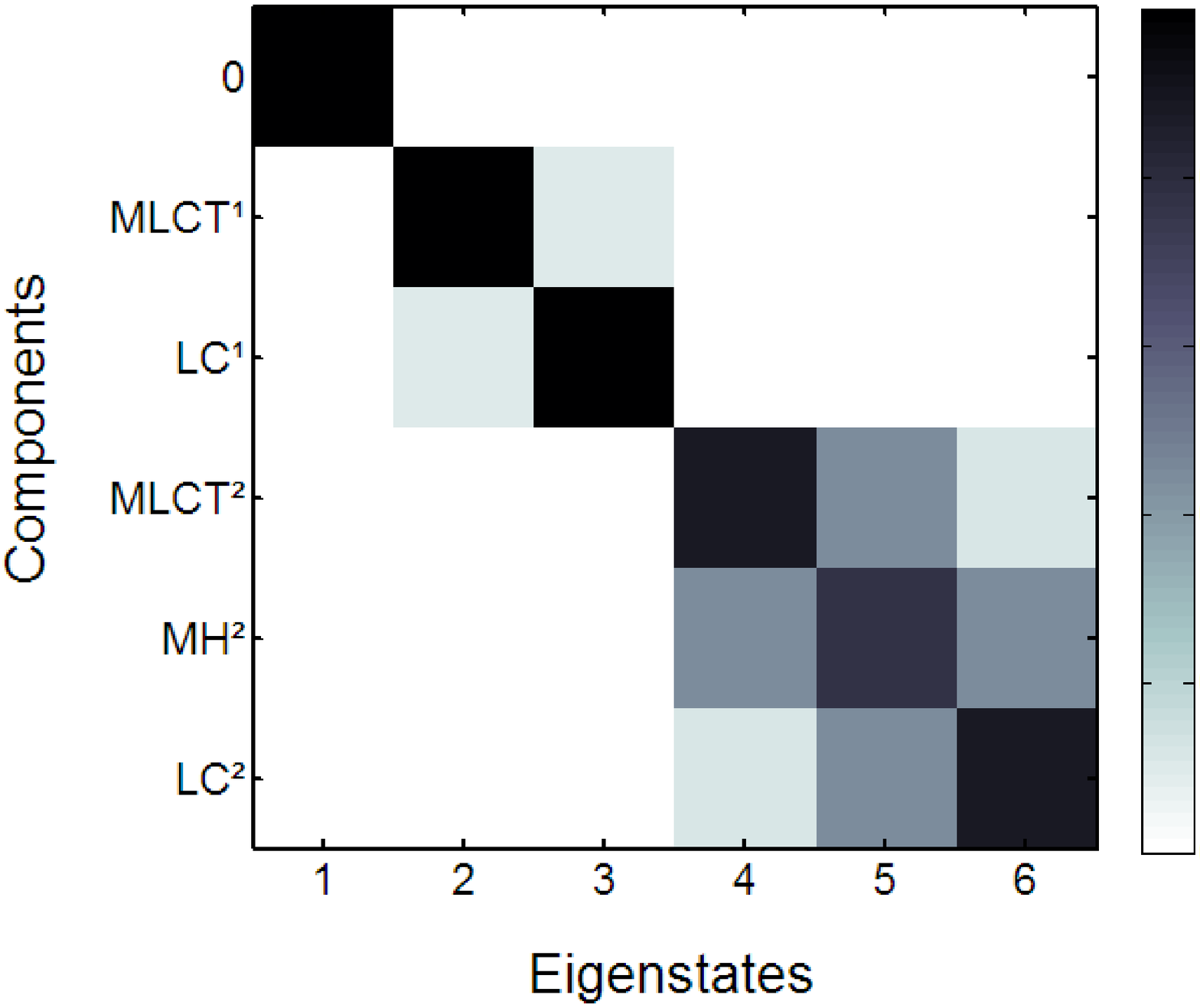}
		\caption{This figure shows the singlet eigenstates with $t^{L} = 0$ eV (when $n_L$ is a good quantum number), for comparison with Fig. 2, showing that dealing with blocks of constant $n_L$ is a valid approximation.}
  \label{fig:robust_eigenstates_tl_0}
\end{figure}

\begin{figure}
\centering
		\includegraphics[width=0.9\columnwidth]{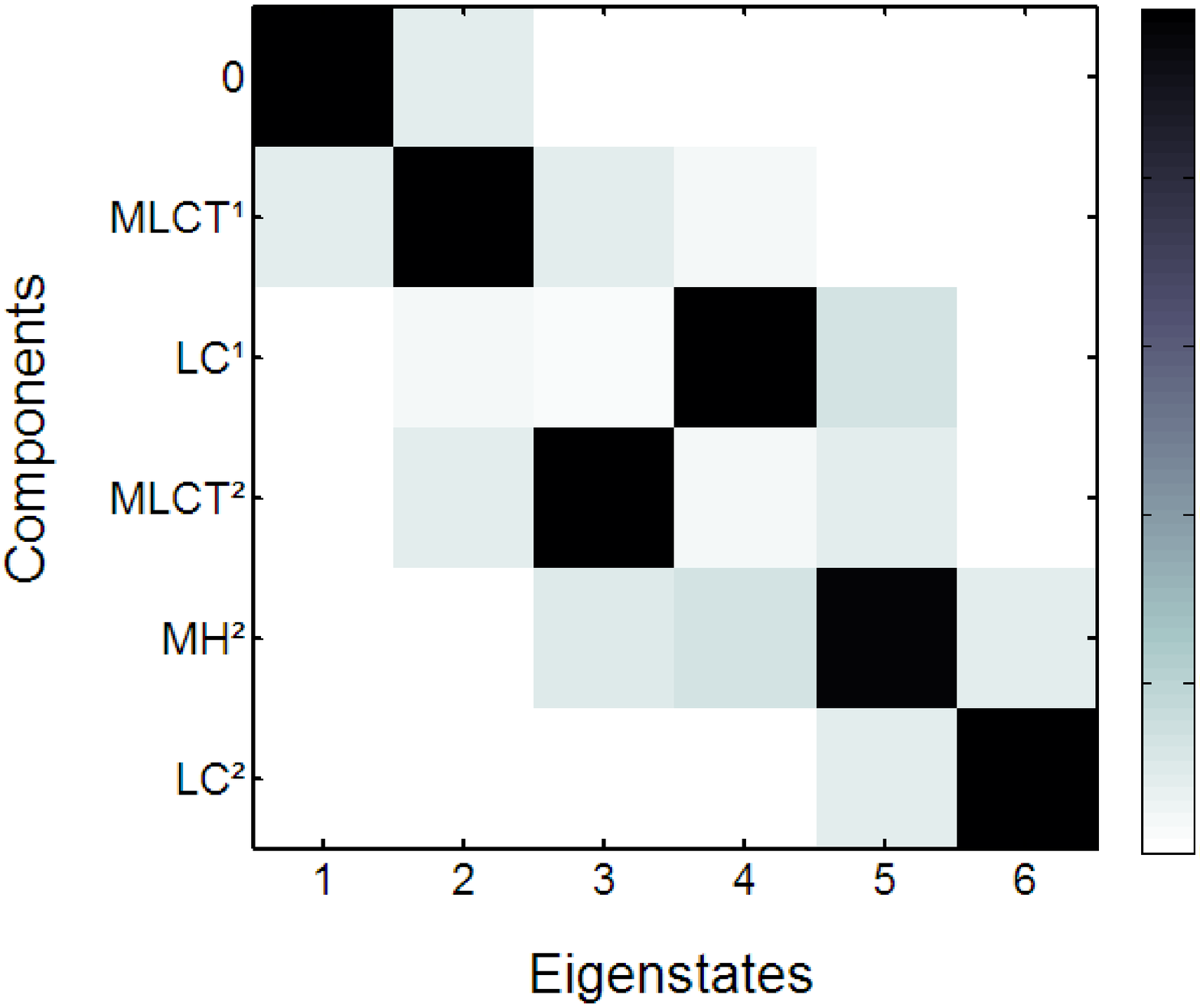}
	\caption{This figure shows the singlet eigenstates with $\varepsilon = 1.5$ eV, for comparison with Fig. 2, showing where $\ket{^1 LC^1}$ and $\ket{^1 MLCT^2}$ have changed order in energy.}
	\label{fig:robust_eigenstates_epsilon_1point5}
\end{figure}

\begin{figure}
	\centering
		\includegraphics[width=0.9\columnwidth]{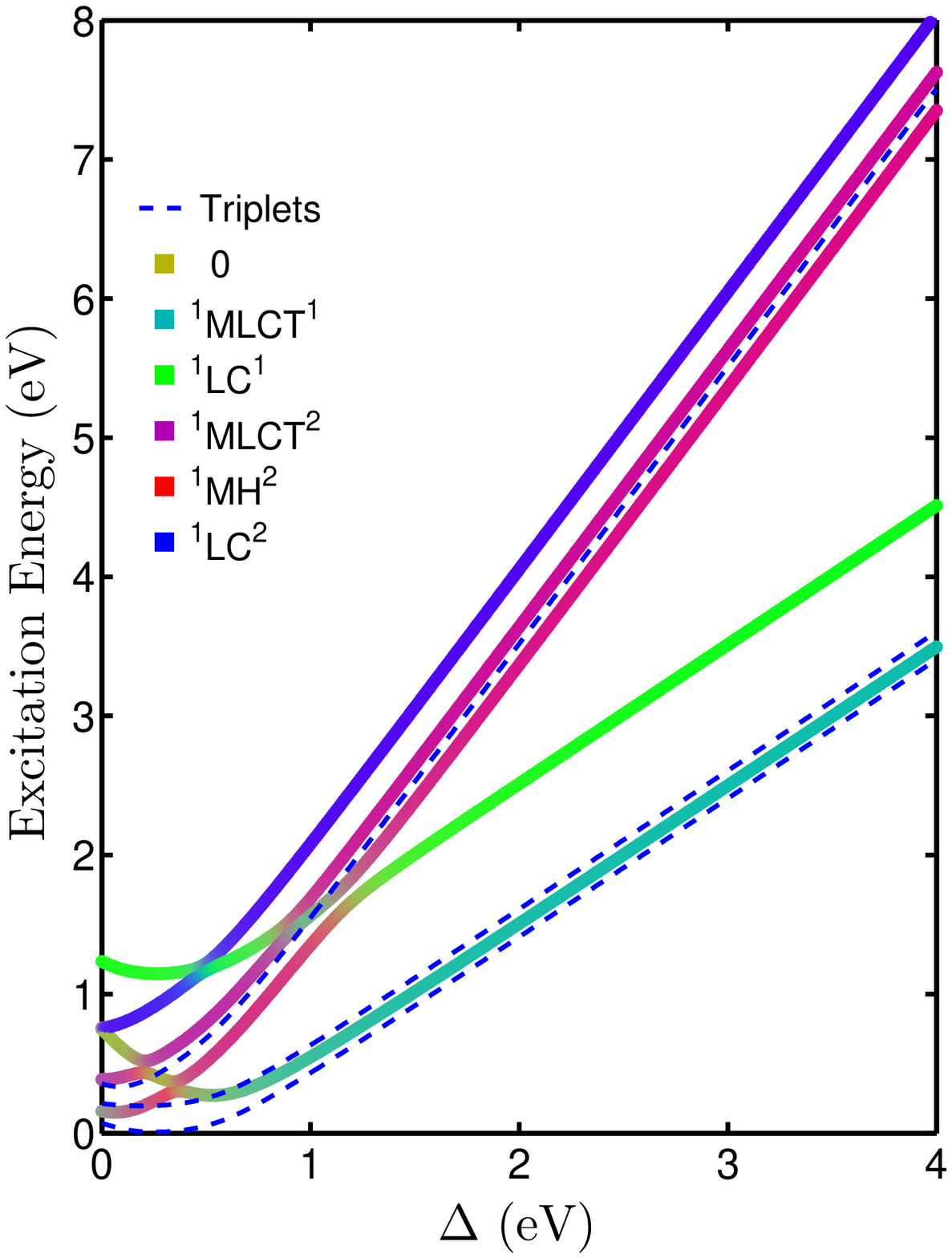}
	\caption{Singlet excitation energies as in Fig. 3 using $V = U - 1/4$ eV. The similarity between this figure and Fig. 3 shows that the approximation $V=U$ does not have any strong effect on the character of the eigenstates.}
	\label{fig:supp_singlet_spec_V_2point75}
\end{figure}
\begin{figure}
	\centering
		\includegraphics[width=0.9\columnwidth]{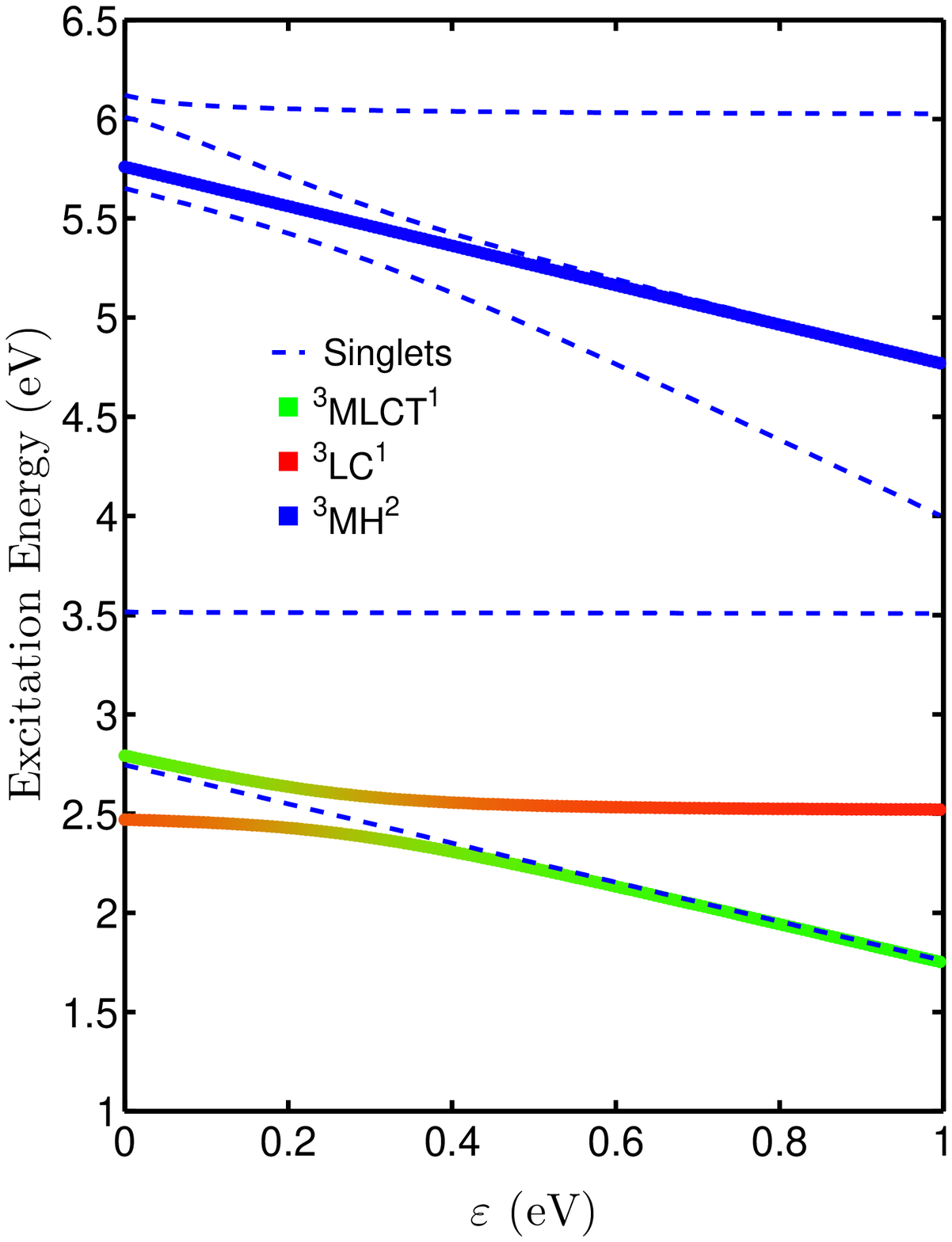}
	\caption{Triplet excitation energies as in Fig. 4 using $V = U - 1/4$ eV. The similarity between this figure and Fig. 4 shows that the approximation $V=U$ does not have any strong effect on the character of the eigenstates.}
	\label{fig:supp_triplet_spec_V_2point75}
\end{figure}

\begin{figure}
	\centering
		\includegraphics[width=0.9\columnwidth]{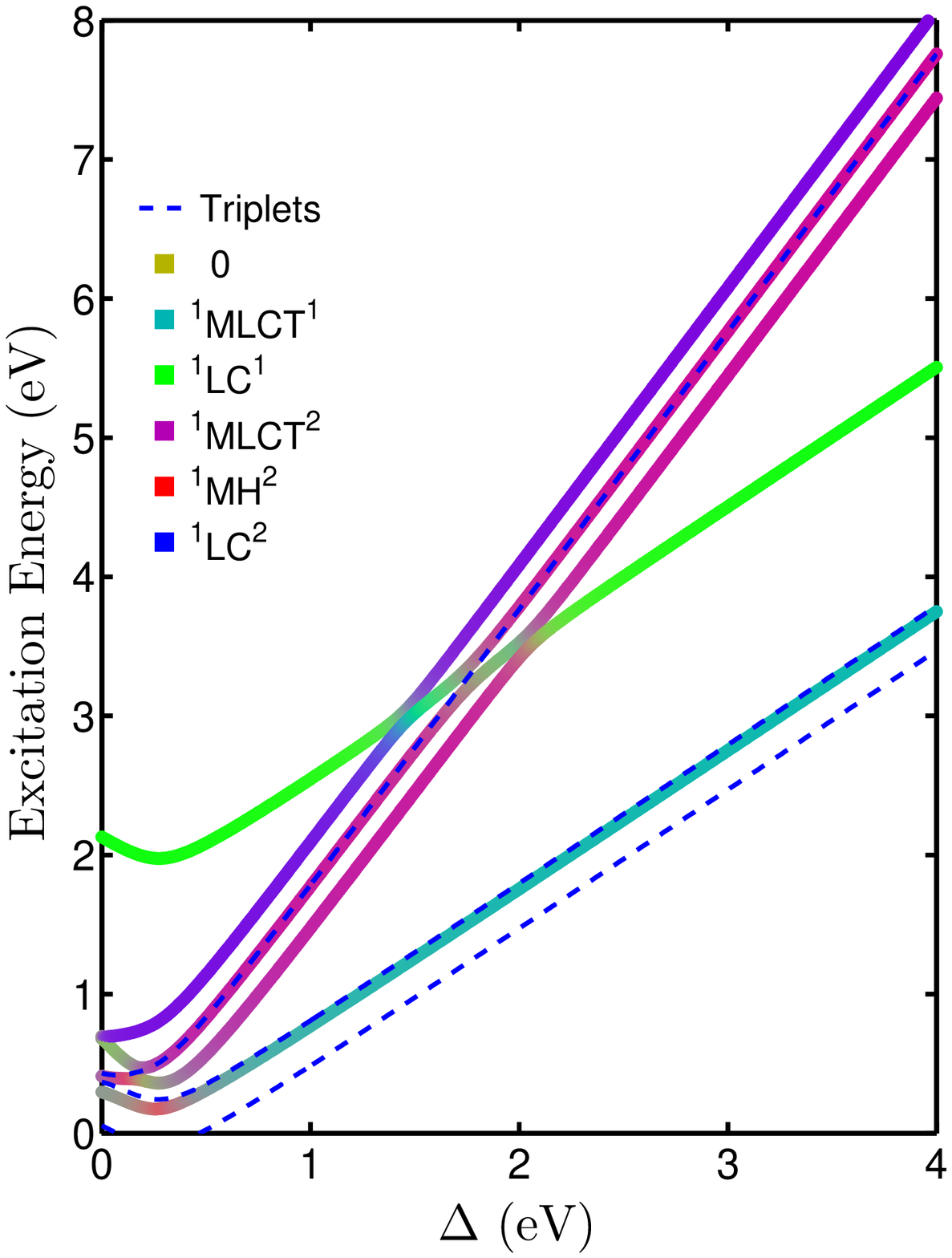}
	\caption{Singlet excitation energies as in Fig. 3 using $J = 2$ eV. The similarity between this figure and Fig. 3 shows that the exact value of $J$ does not have any strong effect on the character of the singlet eigenstates, but does change the value of $\varepsilon$ at which the lowest triplet state changes character.}
	\label{fig:supp_singlet_spec_J_2}
\end{figure}
\begin{figure}
	\centering
		\includegraphics[width=0.9\columnwidth]{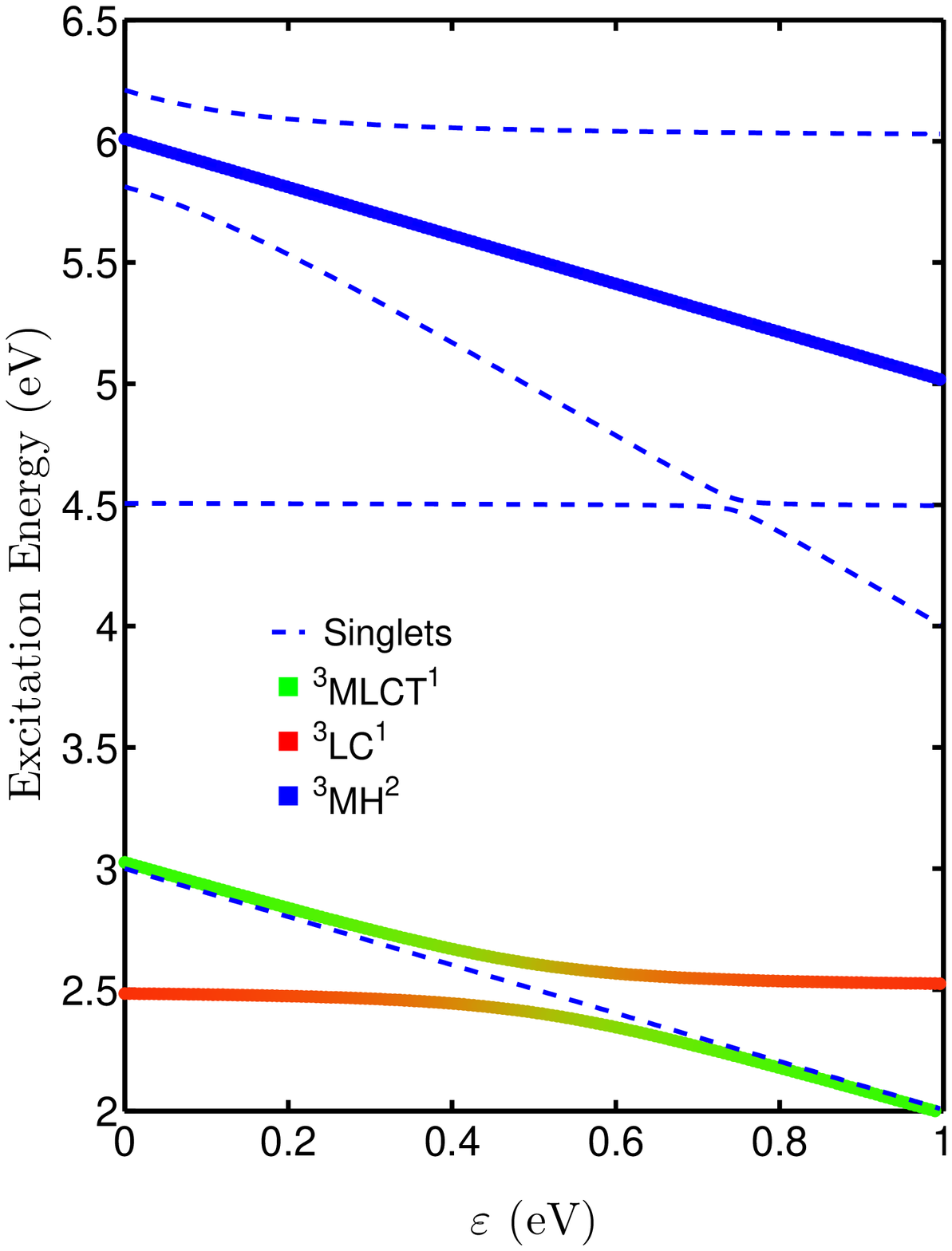}
	\caption{Triplet excitation energies as in Fig. 4 using $J = 2$ eV. The similarity between this figure and Fig. 4 shows that the exact value of $J$ does not have any strong effect on the character of the singlet eigenstates, but does change the value of $\varepsilon$ at which the lowest triplet state changes character.}
	\label{fig:supp_triplet_spec_J_2}
\end{figure}

\begin{figure}
	\centering
		\includegraphics[width=0.9\columnwidth]{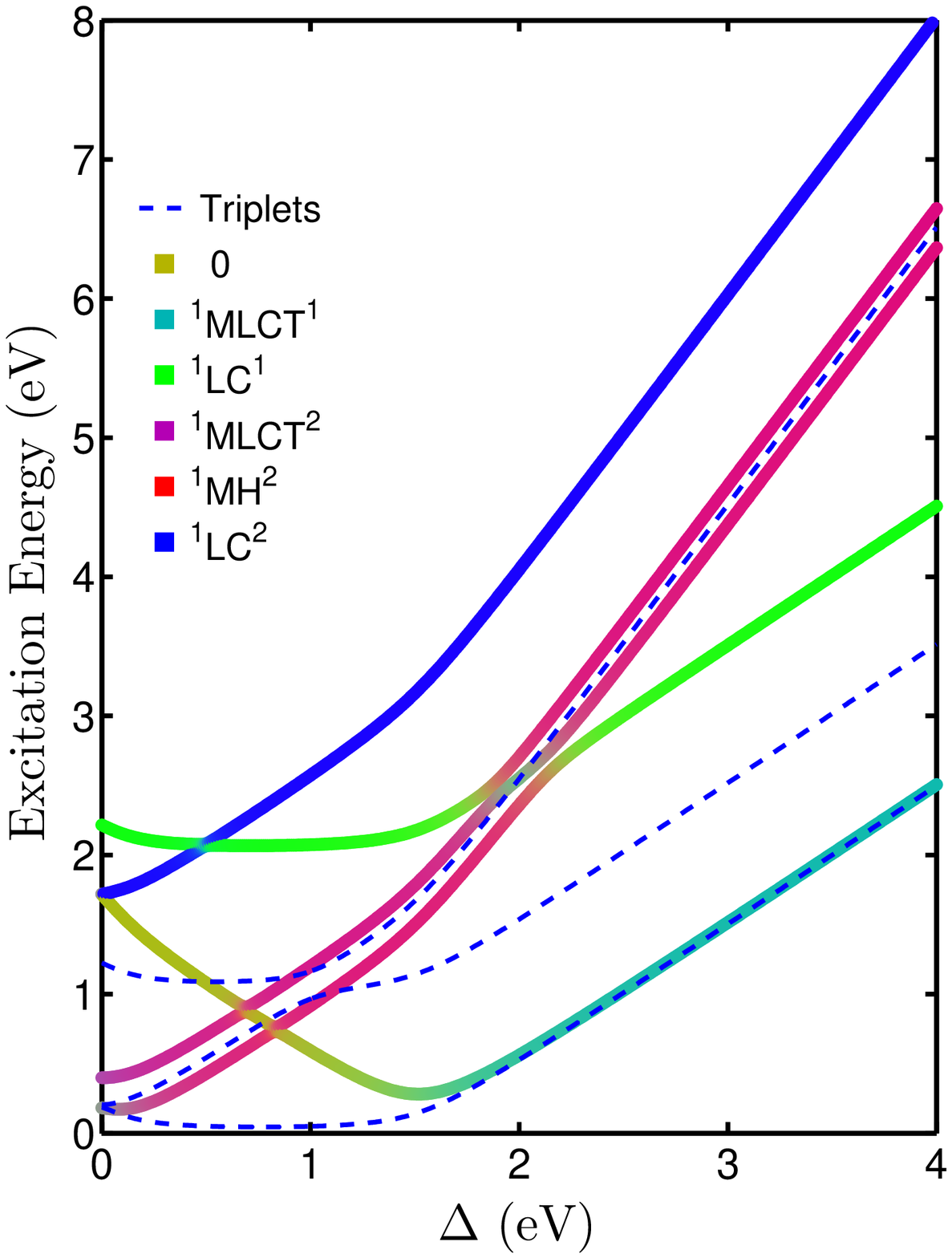} 
	\caption{Singlet excitation energies as in Fig. 3 using $U_M = 4$ eV. The similarity between this figure and Fig. 3 shows that the approximation $U_M=U_H$ does not have any strong effect on the character of the lowest singlet state for the expected range of $\Delta > 3$ eV (determined in the Appendix).}
	\label{fig:supp_singletspec_UM_4}
\end{figure}
\begin{figure}
	\centering
		\includegraphics[width=0.9\columnwidth]{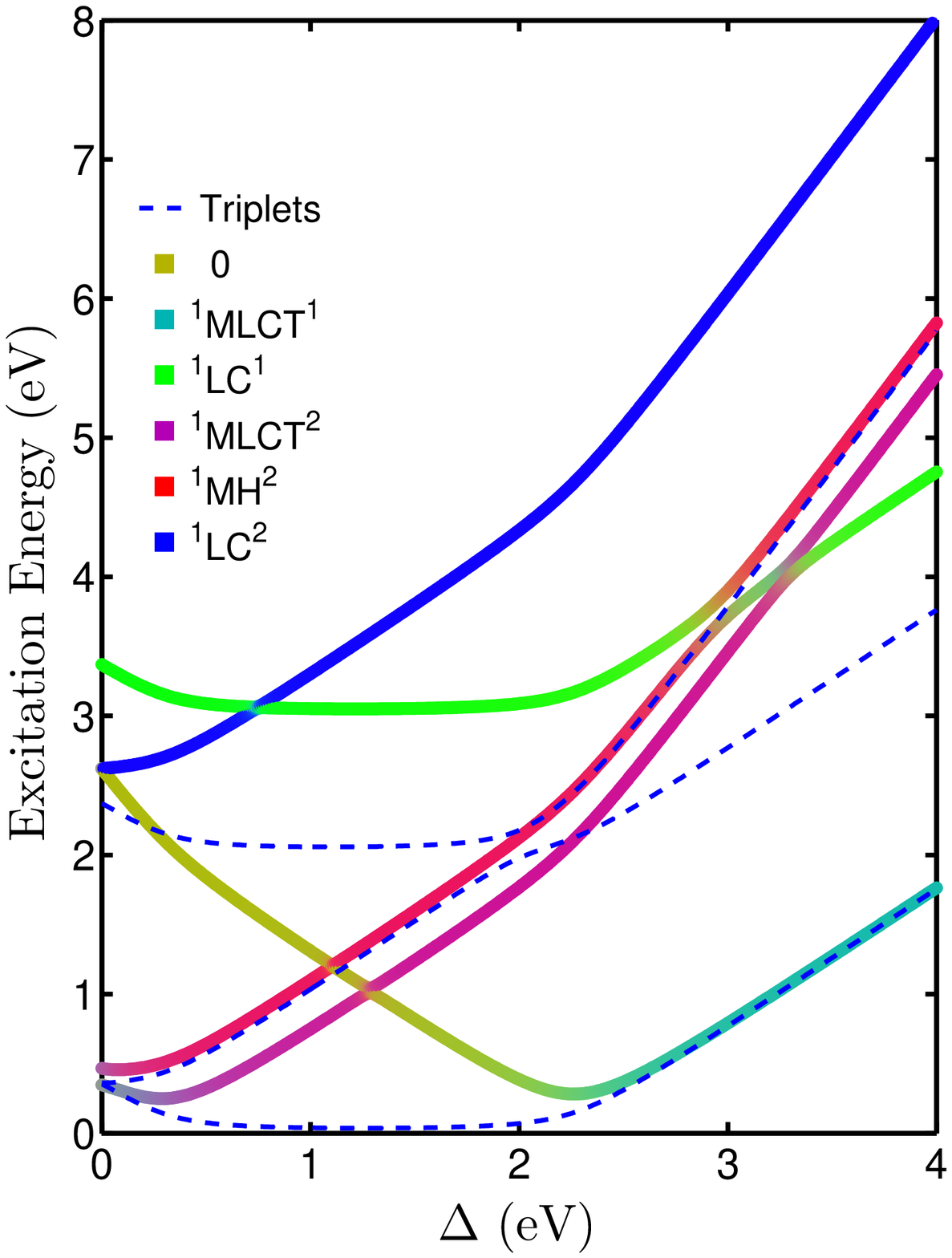}
	\caption{Singlet excitation energies as in Fig. 3 using $U_M = 5$ eV. The similarity between this figure and Fig. 3 shows that the approximation $U_M=U_H$ does not have any strong effect on the character of the lowest singlet state for the expected range of $\Delta > 3$ eV (determined in the Appendix).}
	\label{fig:supp_singletspec_UM_5}
\end{figure}

\begin{figure}
	\centering
		\includegraphics[width=0.9\columnwidth]{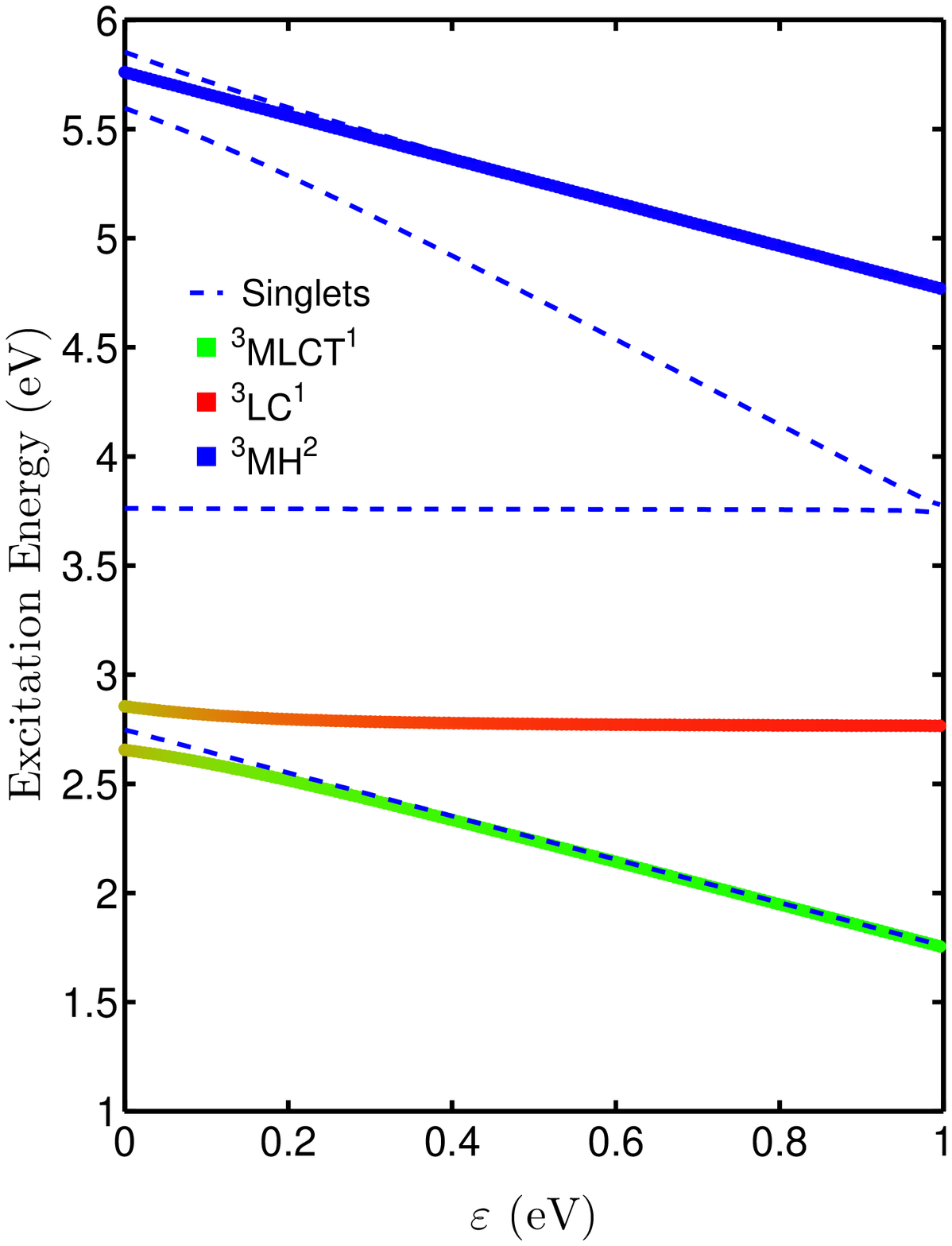} 
	\caption{Triplet excitation energies as in Fig. 4 using $U_M = 3.25$ eV. This figure and Fig. S14 show that increasing $U_M-U_H$ shifts the $\ket{^3 MLCT^1}$-$\ket{^3 LC^1}$ degeneracy condition to lower values of $\varepsilon$ (equivalent to decreasing $J$).}
	\label{fig:supp_tripletspec_UM_3point25}
\end{figure}
\begin{figure}
	\centering
		\includegraphics[width=0.9\columnwidth]{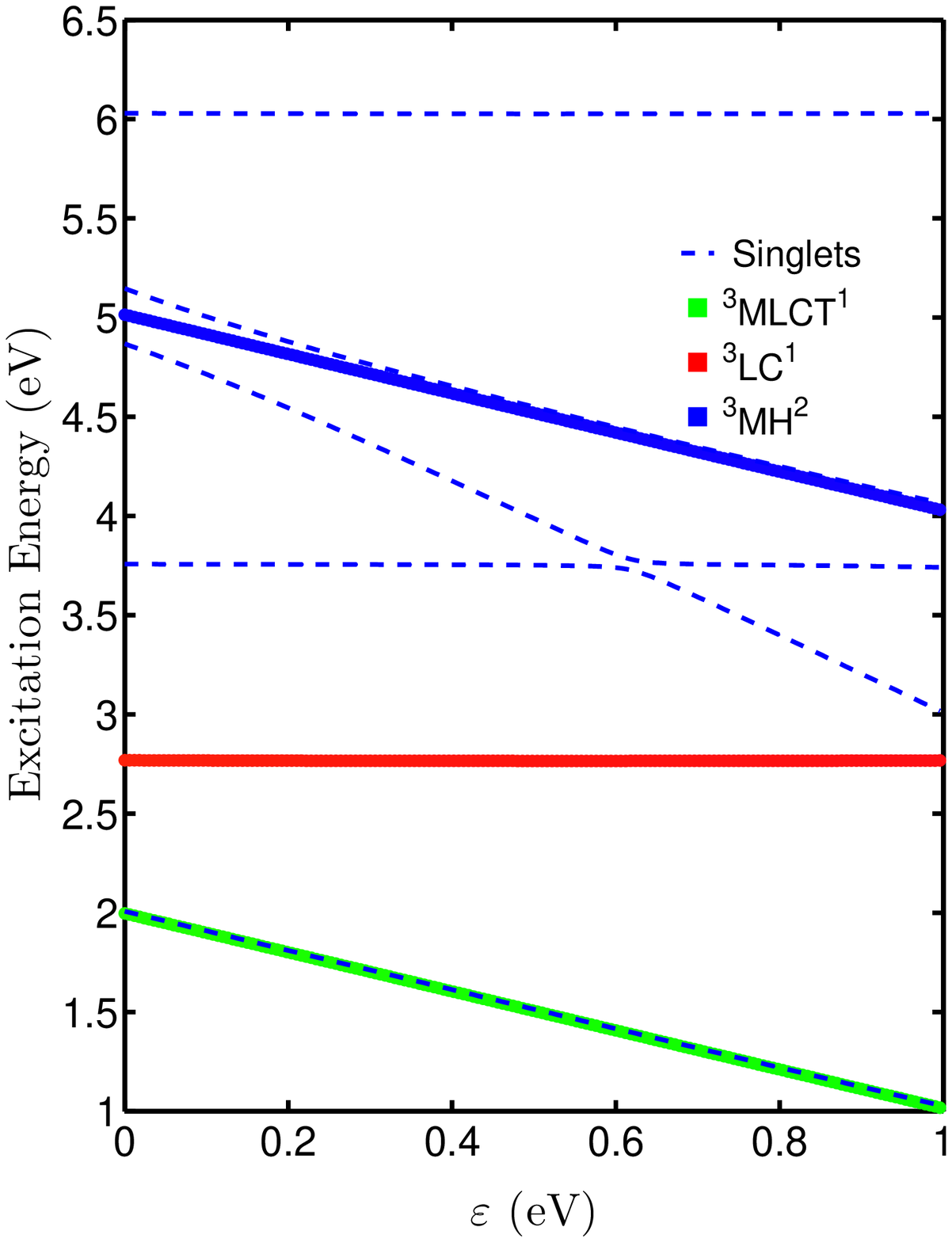}
	\caption{Triplet excitation energies as in Fig. 4 using $U_M = 4$ eV. This figure and Fig. S13 show that increasing $U_M-U_H$ shifts the $\ket{^3 MLCT^1}$-$\ket{^3 LC^1}$ degeneracy condition to lower values of $\varepsilon$ (equivalent to decreasing $J$).}
	\label{fig:supp_tripletspec_UM_4}
\end{figure}

\begin{figure}
	\centering
		\includegraphics[width=0.9\columnwidth]{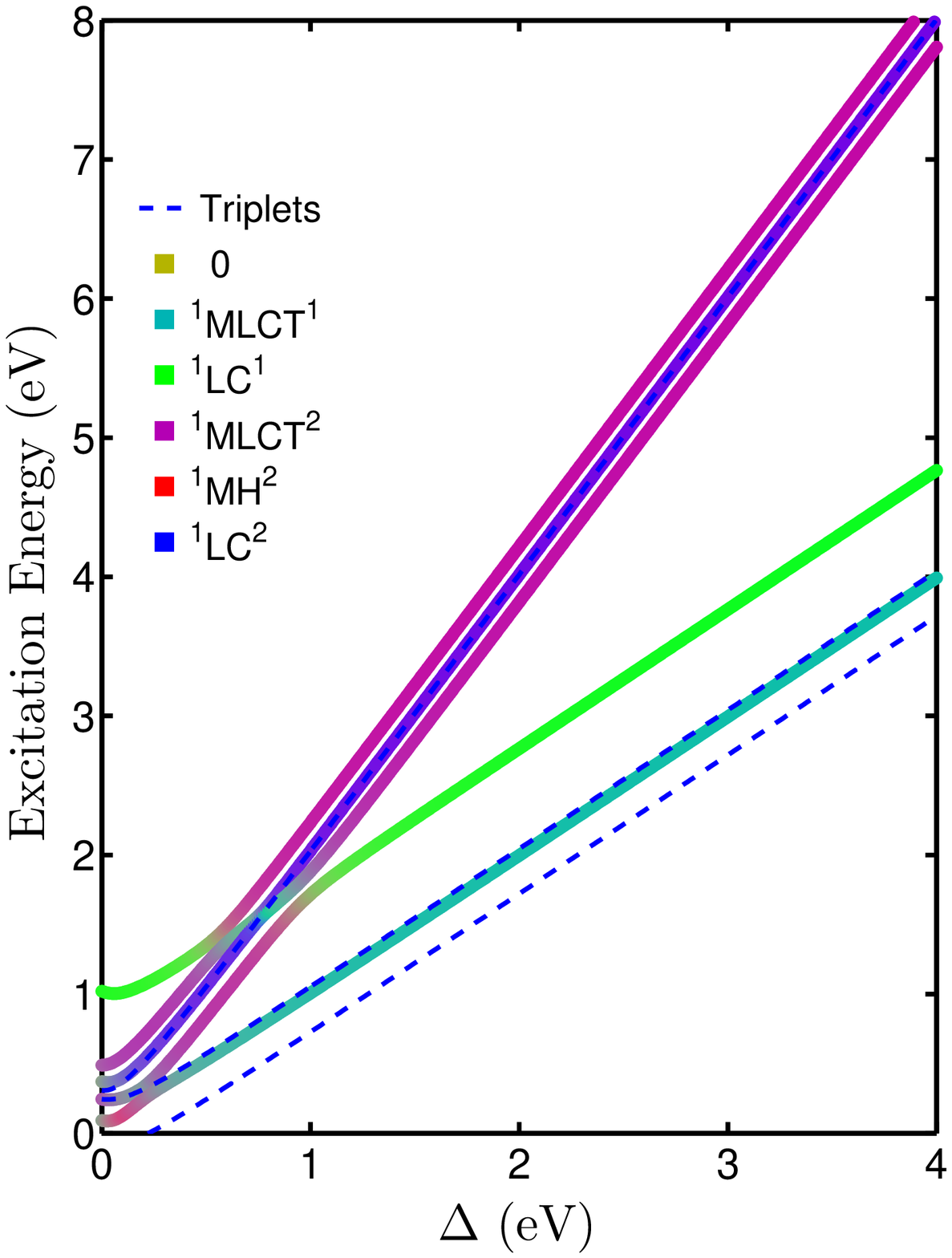} 
	\caption{Singlet excitation energies as in Fig. 3 using $\varepsilon = 0$ eV. The similarity between this figure and Fig. 3 shows that the value of $\varepsilon$ has no important physical effect on the lowest excited singlet states for the expected range of $\Delta > 3$ eV (determined in the Appendix).}
	\label{fig:supp_singletspec_epsilon_0}
\end{figure}
\begin{figure}
	\centering
		\includegraphics[width=0.9\columnwidth]{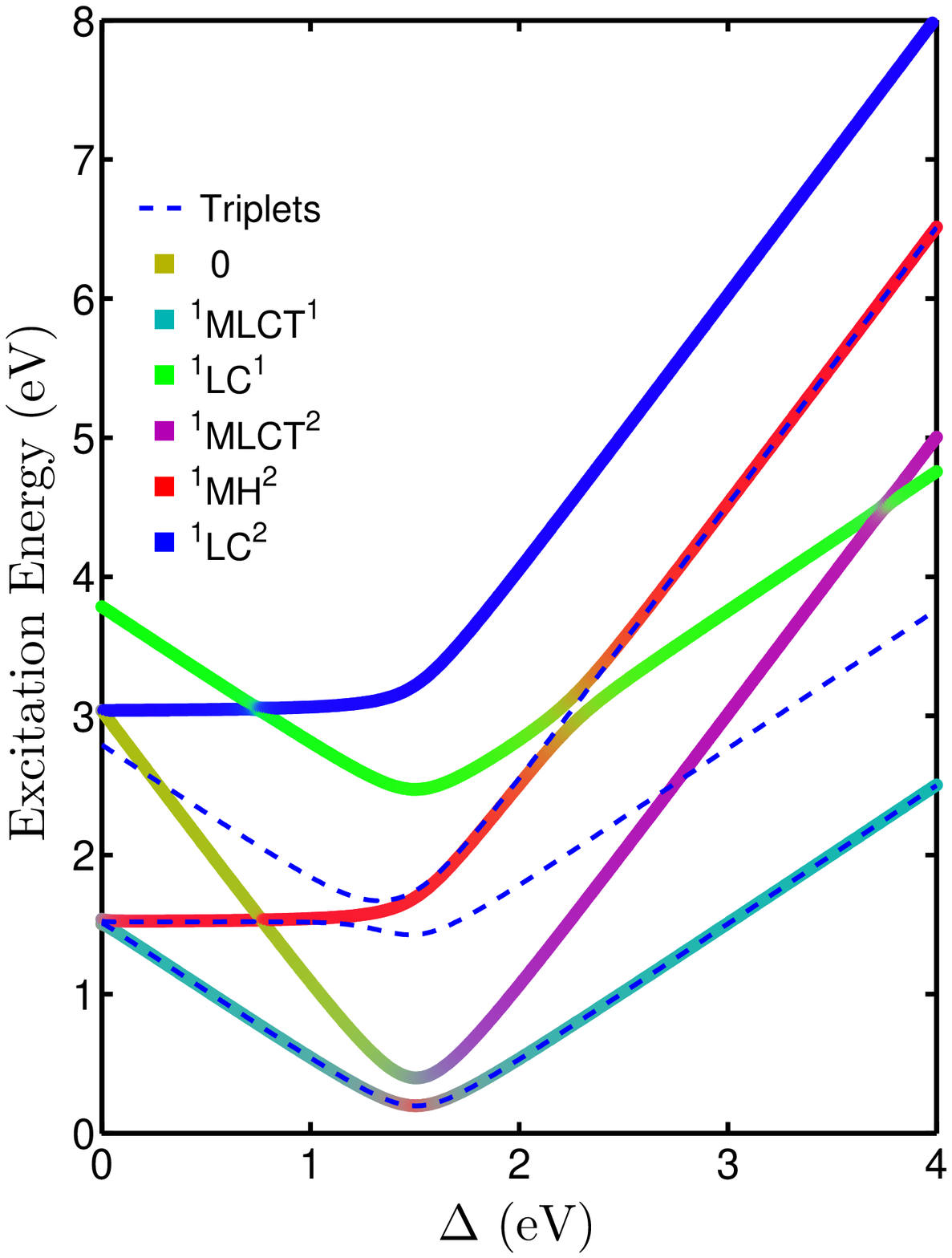}
	\caption{Singlet excitation energies as in Fig. 3 using $\varepsilon = 1.5$ eV. The similarity between this figure and Fig. 3 shows that the value of $\varepsilon$ has no important physical effect on the lowest excited singlet states for the expected range of $\Delta > 3$ eV (determined in the Appendix).}
	\label{fig:supp_singletspec_epsilon_1point5}
\end{figure}

\begin{figure}
	\centering
		\includegraphics[width=0.9\columnwidth]{supp_singletspec_UM_4} \vspace{1 cm} \\
		\includegraphics[width=0.9\columnwidth]{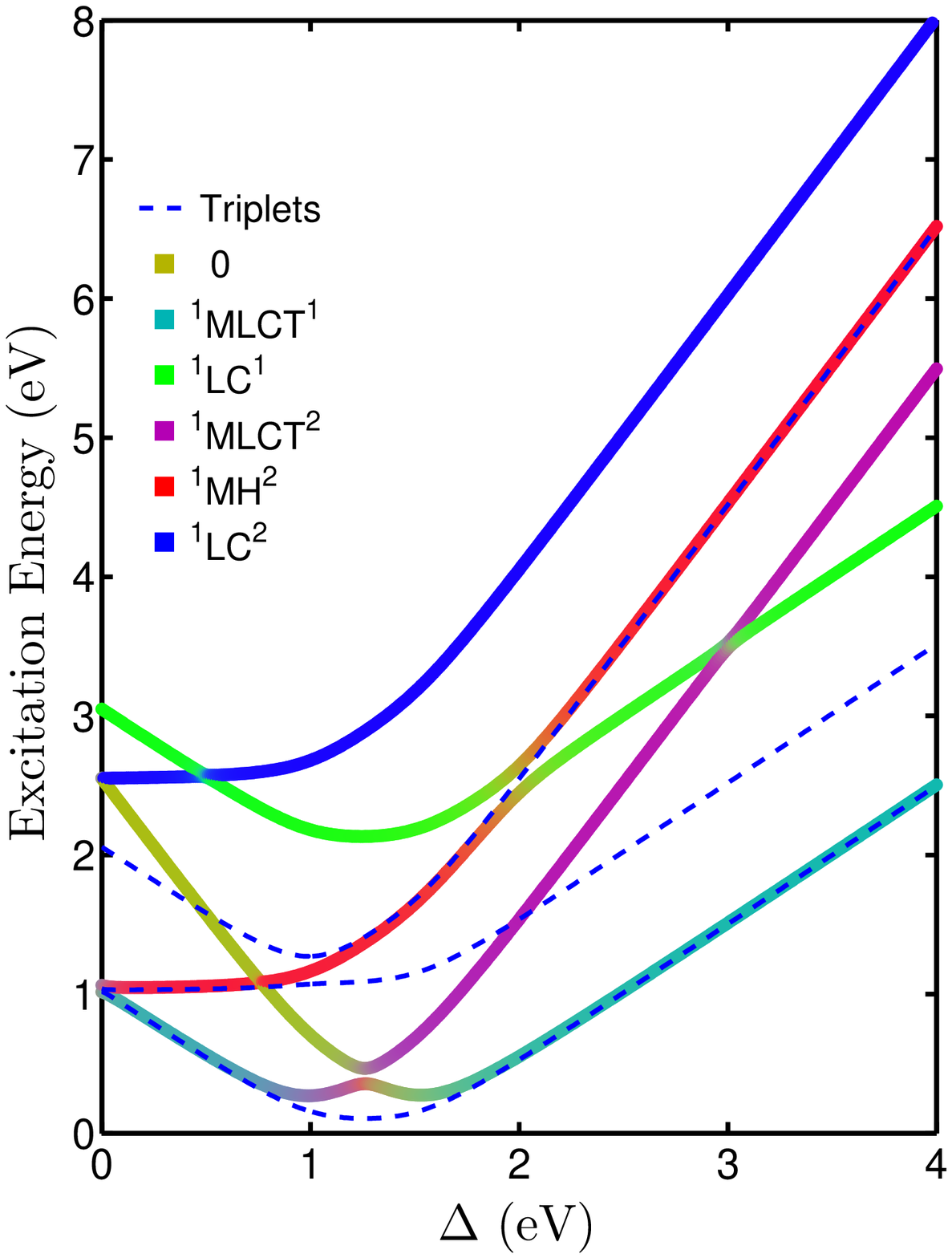}
	\caption{Singlet excitation energies as in Fig. 3 using $U_M = 4$ eV (top) and $\varepsilon = 1.25$ eV (bottom). Both are equivalent to increasing $\varepsilon^*$ by 1 eV. 
	For small $\Delta$ values, these are very different. However, in the range of values expected for $\Delta$ found in the Appendix ($\Delta > 3$ eV), the lowest excited singlet and triplet states are the same. This shows that the lowest excited states are only controlled by $\varepsilon^*$. If we are interested in higher excited states, or if we were to vary $\varepsilon^*$ more, then states with $n_L \neq 1$ would be important in the lowest excited states and varying $U_M$ would no longer be equivalent to varying $\varepsilon$.}
	\label{fig:supp_singletspec_epsilonstar}
\end{figure}